\documentclass[useAMS,usenatbib]{mn2e}

\usepackage[fleqn]{amsmath}
\usepackage{epsfig}
\usepackage{graphicx}
\usepackage{ulem}
\usepackage{color}

\DeclareMathAlphabet{\mathpzc}{OT1}{pzc}{m}{it}

\newcommand\bz{\bar{z}}
\newcommand\bzc{\bar{z}_{\rm c}}

\newcommand\I{\rm{i}}
\newcommand\E{\rm{e}}

\def\beq{\begin{equation}} 
\def\eeq{\end{equation}} 
\def\Dl{D_{\rm l}}
\def\Ds{D_{\rm s}}
\def\Dls{D_{\rm ls}}

\def\apj{ApJ}%
\def\apjl{ApJ}%
\def\aap{A\&A}%
\def\mnras{MNRAS}%
\title[Degeneracies in triple gravitational microlensing]
      {Degeneracies in triple gravitational microlensing}
\author[Song, Mao \& An] {Ying-Yi Song$^{1, 2}$, Shude Mao$^{1, 3}$, Jin H. An$^1$ \\
$^{1}$ National Astronomical Observatories, Chinese Academy of
Sciences, 20A Datun Road, Chaoyang District, Beijing 100012, China \\
$^{2}$ University of Chinese Academy of
Sciences, 19A Yuquan Road, Shijingshan District, Beijing 100049, China \\
$^{3}$ Jodrell Bank Centre for Astrophysics, The University of
Manchester, Alan Turing Building, Manchester M13 9PL, UK \\
}
\begin{document}
\include{journaldefs}
\date{Accepted ...... Received ...... ; in original form......   }

\pagerange{\pageref{firstpage}--\pageref{lastpage}} \pubyear{2013}
\maketitle
\label{firstpage}

\begin{abstract}
We study microlensing light curves by a triple lens, in particular, by a primary star plus two planets. A four-fold degeneracy is confirmed in the light curves, similar to the close and wide degeneracy found in a double lens. Furthermore, we derive a set of equations for triple-lens in the external shear approximation. By using these external shear equations, we identify two kinds of continuous degeneracies which may confused double and triple lenses, i.e. the continuous external shear degeneracy among triple-lens systems and the double-triple lens degeneracy. These degeneracies are particularly important in high magnification events, and thus some caution needs to be applied when one infers the fraction of stars hosting multiple planets from microlensing. We study the dependence of the degeneracies on the lensing parameters (e.g., source trajectory) and give recipes about how the degeneracies should be explored with real data.
\end{abstract}

\maketitle


\section{Introduction}
\label{sec:introduction}

The single point lens equation can be analytically solved \citep{Pac86}. The lens equation for a double lens becomes considerably more complex, and is no longer analytical \citep{SW86, MP91}. In this case, when the source is far away from the caustics, there are always three images; when the source is inside the caustics, the number of images increases by two. It is analytically known that for five image configurations, the minimum total magnification is 3 \citep{WM95, Rhi97}. 

The light curve of a double lens can be diverse, depending on the lens parameters and source trajectory. For extreme mass ratios, there is a well-known degeneracy between close and wide separation binaries which yields essentially identical light curves \citep{Dom99}. Furthermore, planetary and stellar double lens light curves can also mimic each other \citep{Cho12}. Many of these degeneracies can be (partially) broken with accurate photometry and cadence, or with additional information from parallax (e.g., \citealt{Gou92, Smi03}) or finite source size effects (\citealt{WM94, Gou94, Nem94}).

Nevertheless an analytical understanding of the degeneracy in the lens equation is helpful in searching the parameter space. Historically, a wrong solution has been picked in the presence of degeneracy for the parallax microlensing event MACHO-LMC-5, which was later corrected with analytical insight \citep{Smi03, Gou04, Dra04}.

\cite{Gau98} pointed out that for some geometries, the magnification pattern and resulting light curves from multiple planets are qualitatively degenerate with those from single-planet lensing without providing mathematical explanations. \cite{Boz00} examined the caustics of multiple lenses in two extreme cases, i.e., the separations between each two lenses are either very large with respect to their Einstein radii, or very small compared to the Enstein radius of the total mass. He also demonstrated a principle of duality between planets external and internal to the Einstein ring, which turned out to be the close-wide degeneracy for multiple stars.

The recent discovery of two double-planet systems \citep{Gau08, Han13} illustrates a need for exploring further the degeneracy for multiple ($N \ge 3$) lenses or new degeneracies yet to be found. Due to the greater number of parameters in triple lensing, the search of the parameter space is even more time-consuming, and so analytical guidance becomes even more important. This paper is an attempt to explore this issue. Compared to the previous studies, we explore a few new issues:
(1) We, for the first time, discuss the three-body vs. three-body (\S\ref{sec:triple}) and two-body vs. three-body (\S\ref{sec:double}) degeneracies in great detail (which was mentioned in \citealt{Gau98}).
In particular, we give detailed procedures in \S\ref{sec:degeneracy} and Appendix A how to explore this degeneracy.
(2) We explore the correlation between different parameters (error ellipses) for the first time (Figs. 4 and 5) through concrete examples of light curves.
(3) We also pay more attention to light curves and consider the residual between the degenerate cases to show the strength of the degeneracies.
(4) Technically, the methods we use are somewhat different: we use complex notations and expand the lens equations directly used by Dominik (1999) and An (2005) whose papers are mainly about binary lenses. In contrast, \cite{Boz99, Boz00} mostly used polar coordinates and expand the Jacobian determinant to investigate the caustics of multiple lenses.

The structure of the remaining paper is as follows. In
\S\ref{sec:lensSystem} we present the lens equation and our notations;
in \S\ref{sec:degeneracy} and \S\ref{sec:other} we generalise the degeneracies found in triple lenses; finally, in \S\ref{sec:discussion} we briefly discuss our results.

\section{The triple-lens system}
\label{sec:lensSystem}

We start this section by presenting the $n$-point lens equation, and then introduce the notations we use for later discussions.

\subsection{The lens equation}

In complex notation, the $n$-point lens equation can be written as (\citealt{Witt90})
\begin{equation}  
\label{lensEq}
\zeta = z - f( \bz ), \qquad f( \bz ) =  \sum_{k=1}^N {m_k \over \bz - \bz_k}
\end{equation}
where $\zeta$ and $z$ are the source and its lensed image positions,  and $z_k$ and $m_k$ is the position and the mass of the $k$-th lens. Note that, $\bz$ and $\bz_k$ represent complex conjugates of $z$ and $z_k$.

The lens equation describes the mapping from the lens plane onto the source plane. The Jacobian matrix of the mapping is given by
\begin{equation}  \label{Jacobian}
J =
\begin{pmatrix}
 {\partial \zeta \over \partial z} & {\partial \zeta \over \partial \bz } \\
 {\partial \bar{\zeta} \over \partial z } & {\partial \bar{\zeta} \over \partial \bz}
\end{pmatrix}
=
\begin{pmatrix}
 1 & {df \over d\bz } \\
 \overline{df \over d\bz} & 1
\end{pmatrix}
\end{equation}
The determinant of the Jacobian matrix is ${{\rm det} \,J} = 1 - {|{df / d{\bz}}|}^2$. Since gravitational lensing conserves surface brightness, the magnification is simply given by $\mu=|{\rm det}\, J|^{-1}$. Obviously, if ${{\rm det\,} J} = 0$, the magnification $\mu$ is formally infinite. Image positions satisfying this condition form one or more closed ``critical curve(s)'' in the lens plane, which are mapped into ``caustics'' in the source plane. For convenience, we plot all these curves on the same plane in units of the angular Einstein radius (see equation~\ref{rE} in \S\ref{sec:notations}).

\subsection{Notations of Lens Parameters}
\label{sec:notations}

\begin{figure}
\graphicspath{}
\centering
\vspace{-3mm}
\includegraphics[width=.45\textwidth]{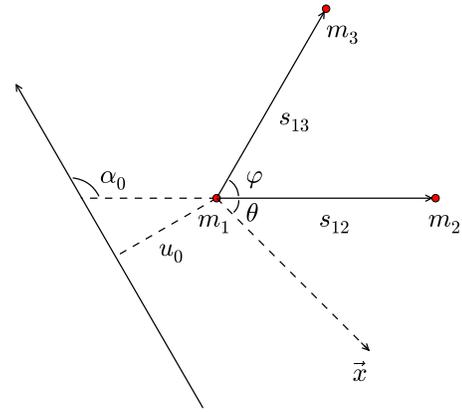}
\caption{Static triple-lens parameters in units of $\theta_{\rm E}$ in two coordinate systems. The three lenses are labelled as $m_1$, $m_2$ and $m_3$. Both coordinate systems are centred on the primary lens $m_1$, and the two separations are labelled as $s_{12}$ and $s_{13}$ with a characteristic angle $\varphi$ between them. In the first coordinate system, $m_1$ and $m_2$ are on the horizontal axis, and the source trajectory is parameterised by two parameters, i.e., the impact parameter $u_0$ and the trajectory angle $\alpha_0$. In the second coordinate system, the $x$-axis in an arbitrary direction is chosen as the horizontal axis and the angle between the $x$-axis and the line connecting $m_1$ and $m_2$ is denoted as the direction angle $\theta$.}
\label{fig:param}
\end{figure}

In this paper, we do not consider blending and finite-source size effect. Moreover, we also do not include microlens parallax and orbital motion effects. So the trajectory of the lens-source relative motion is a straight line, and all the lens systems are static. 

\cite{Gou00} suggested a set of notational conventions for point lens microlensing (see \citealt{Sko11} for further extensions to the double-lens case with orbital motion and even full Keplerian solutions). In their notation the distances to the lens and source are denoted as $\Dl$ and $\Ds$, and the distance between the lens and source are $\Dls$. The angular Einstein radius is given by
\begin{equation}
\label{rE}
 \theta_{\rm E} = \sqrt{{4GM \over c^2} {\Dls \over \Dl \Ds}},
\end{equation}
where $M$ is the mass of the lens.

Throughout this paper, we use angular coordinates which are normalised to $\theta_{\rm E}$ defined above, and the corresponding time is normalised by the Einstein radius crossing time $t_{\rm E}$. The total lens mass $M$ is normalised to unity ($m=1$).

For any lens system, there are three basic parameters $(t_0,\,u_0,\,t_{\rm E})$, where $t_0$ is the time of the closest approach to the lens system ``center'', $u_0$ is the corresponding lens-source projected separation (in units of $\theta_{\rm E}$) at $t_0$, and $t_{\rm E}$ is the Einstein radius crossing time. Normally, we set the origin of the coordinate system at the position of the primary object.

In a static double-lens system, there are two mass components $m=m_1+m_2=1$, where $m_1$ is the primary mass and $m_2$ is the secondary mass ($m_1 \geq m_2$). Three additional parameters are needed to describe the configuration of the double-lens, namely $(m_2,\,s,\,{\alpha}_0)$, or equally $(q,\,s,\,{\alpha}_0)$, where $q = {m_2/m_1}$ is the mass ratio of the binaries, $s$ is the projected separation between the primary and secondary objects (in units of $\theta_{\rm E}$), and ${\alpha}_0$ is the direction of lens-source relative motion with respect to the double-lens axis (primary toward secondary), i.e., the angle from the double-lens axis to the trajectory counterclockwise. 

Similarly, a static triple-lens has six additional parameters (compared to a single lens model)
\begin{equation}
\label{set1}
 (m_2,\,m_3,\,s_{12},\,s_{13},\,\varphi,\,\alpha_0),
\end{equation}
or equally $(q_{21},\,q_{31},\,s_{12},\,s_{13},\,\varphi,\,\alpha_0)$, where $q_{21} = {m_2 / m_1}$, $q_{31} = {m_3 / m_1}$ and $m = m_1 + m_2 + m_3 = 1$. ${\alpha}_0$ is the angle from the $s_{12}$-axis to the trajectory counterclockwise, $\varphi$ is the angle from the $s_{12}$-axis to the $s_{13}$-axis counterclockwise. For later convenience, we also define an angle $\theta$ from the $x$-axis of the chosen coordinate system to the $s_{12}$-axis (measured counterclockwise). Henceforth, we call $\alpha_0$ the ``trajectory angle'', $\varphi$ the ``characteristic angle'' and $\theta$ the ``direction angle''. Fig.~\ref{fig:param} illustrates all the parameters in a static triple-lens system. 

Specially, in the absence of parallax effects, a static triple-lens system has an obvious discrete degeneracy,
\begin{equation}
 (u_0,\,\alpha_0,\,\varphi) \rightarrow -(u_0,\,\alpha_0,\,\varphi).
\end{equation}
It also indicates an axial symmetry which can reduce the range of $\varphi$ from $[0^\circ, 360^\circ]$ to $[0^\circ, 180^\circ]$. 

\section{Degeneracy in triple lensing}
\label{sec:degeneracy}

In this section, we first explore the close/wide degeneracy in triple lensing with two planets, and then a continuous degeneracy arising in the external shear approximation. These theories are mainly suitable to the central caustics, which are defined as the caustics in the vicinity of the primary object.

\subsection{The planetary close/wide degeneracy}

\begin{figure*}
\graphicspath{}
\centering
\vspace{-3mm}
\includegraphics[width=.45\textwidth]{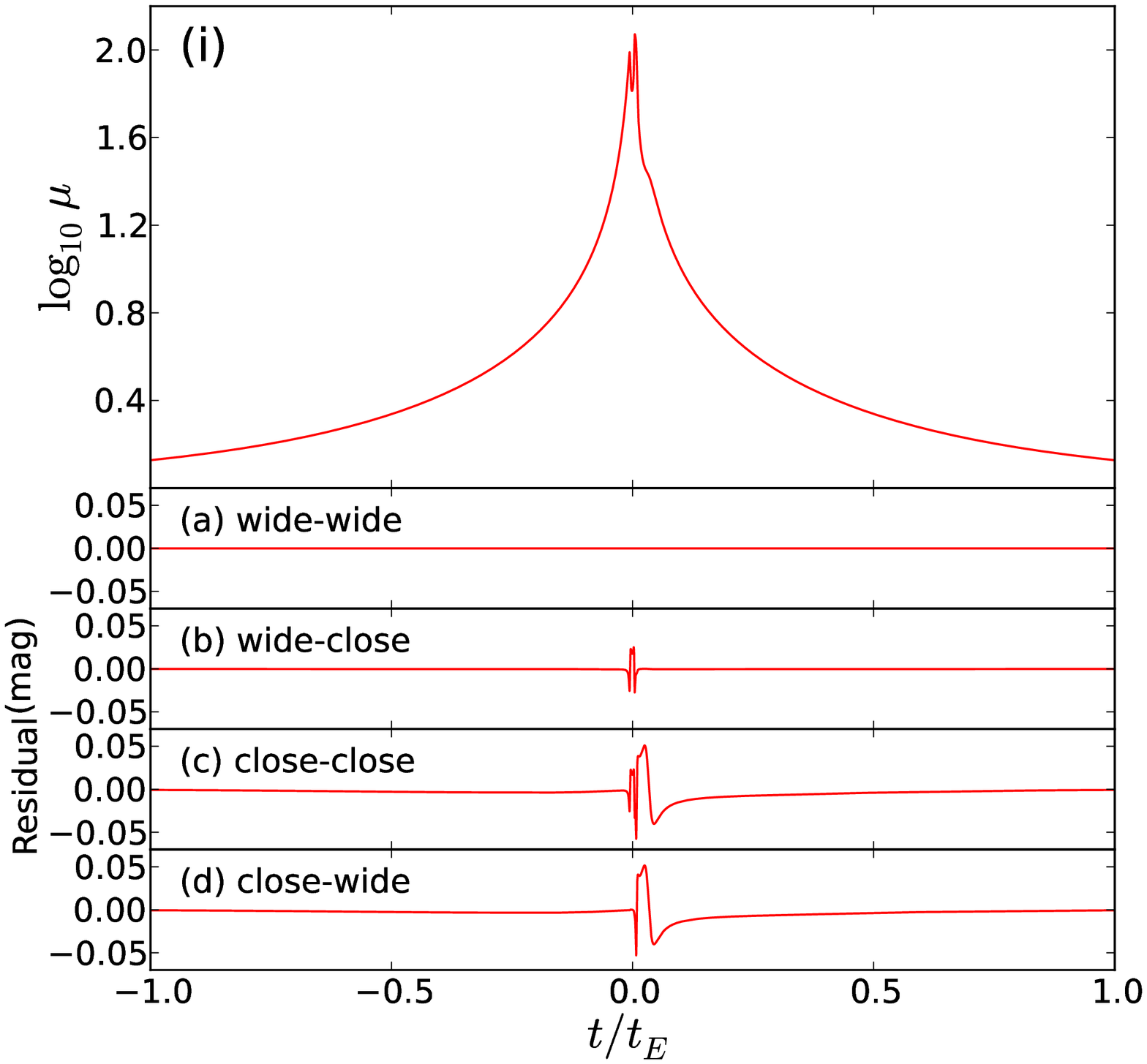}
\includegraphics[width=.45\textwidth]{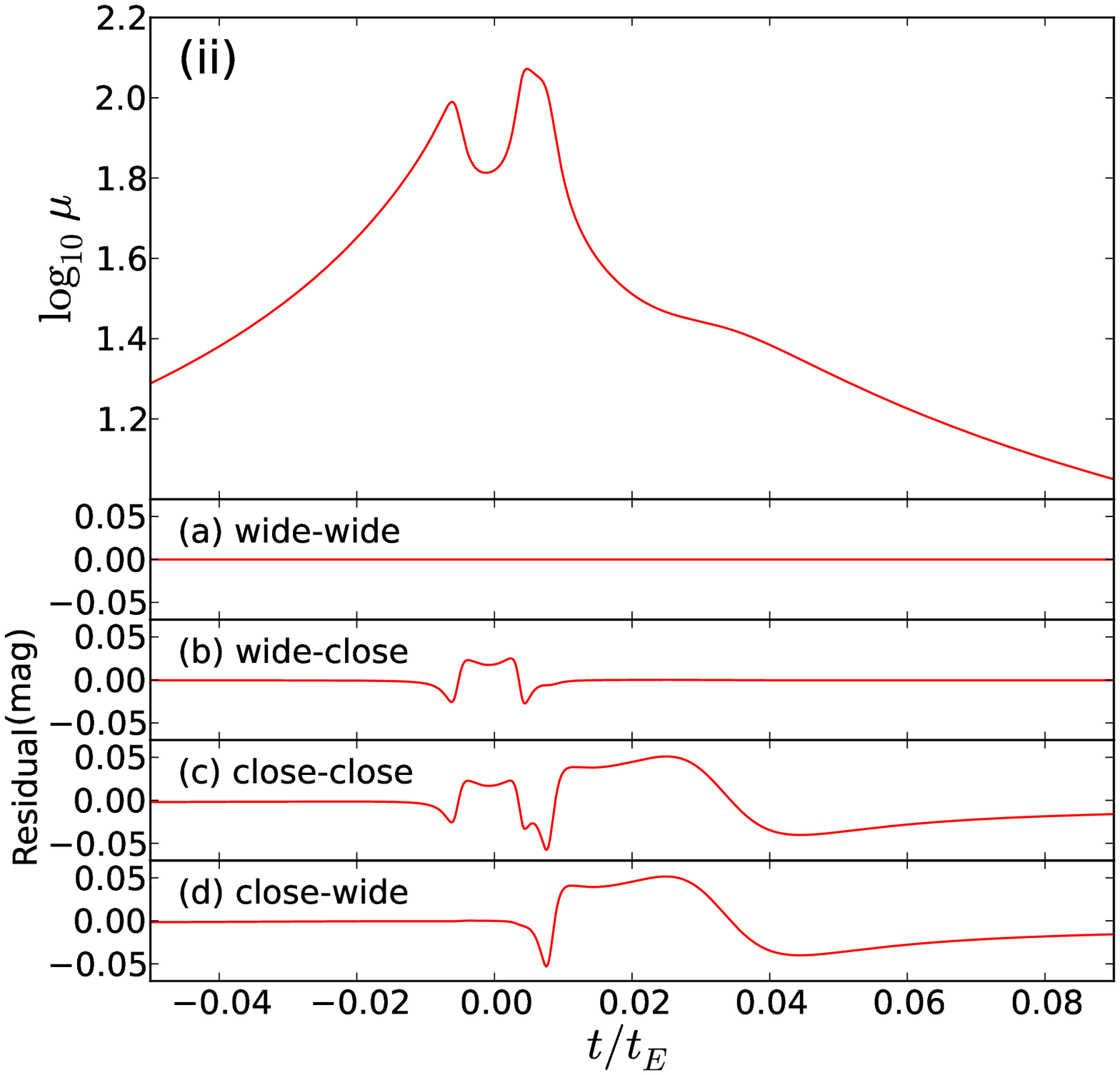}
\includegraphics[width=.45\textwidth]{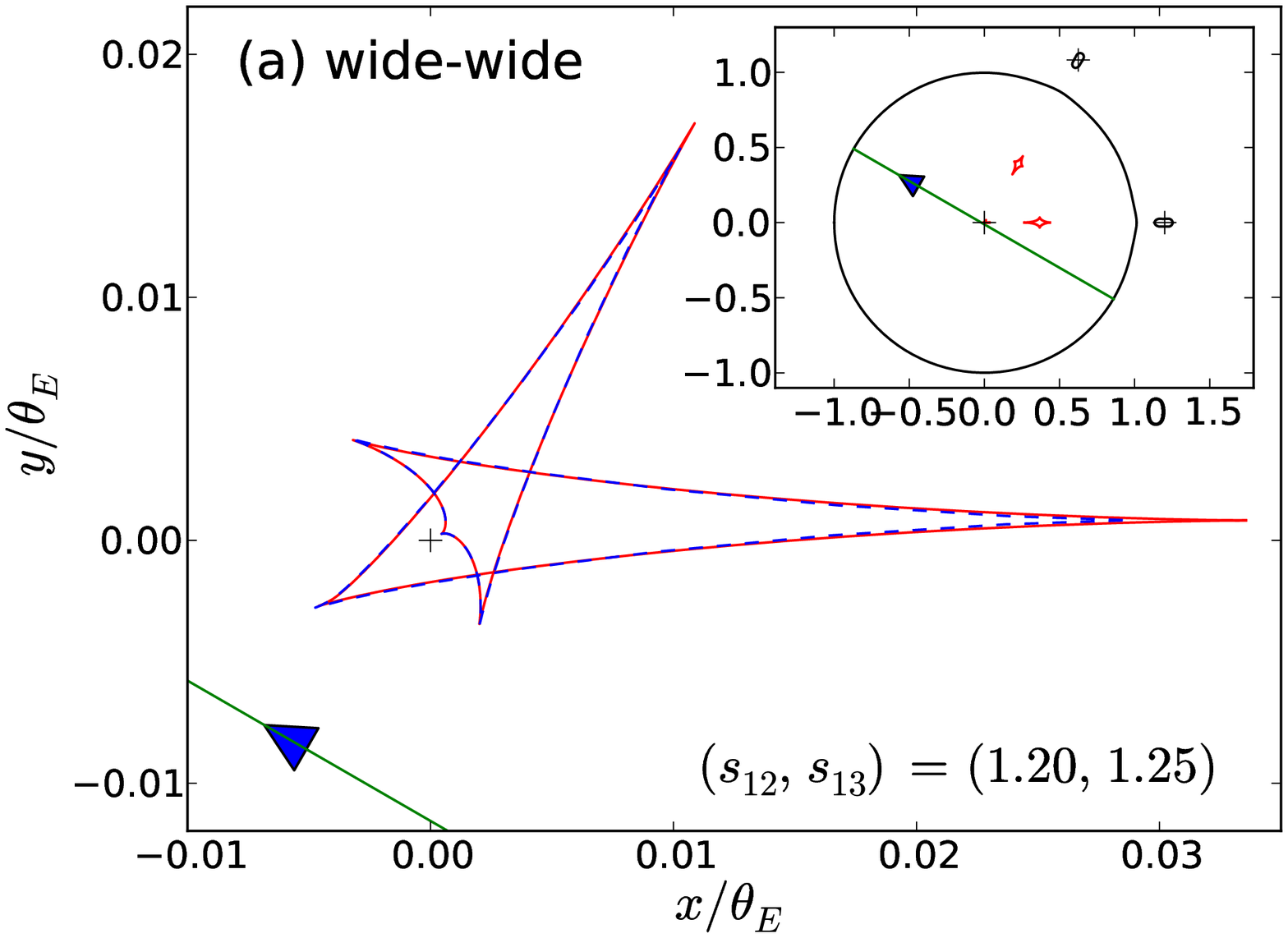}
\includegraphics[width=.45\textwidth]{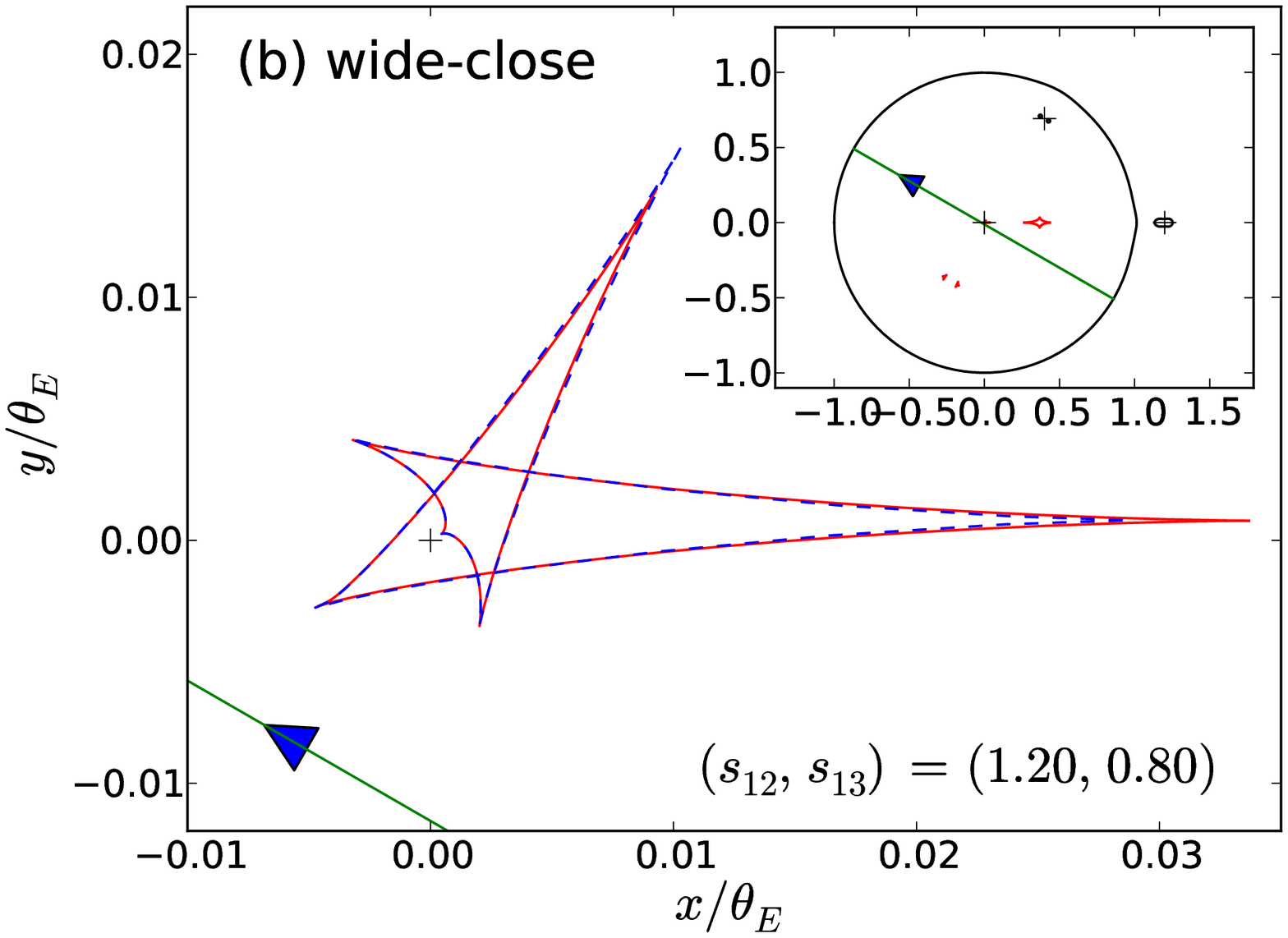}
\includegraphics[width=.45\textwidth]{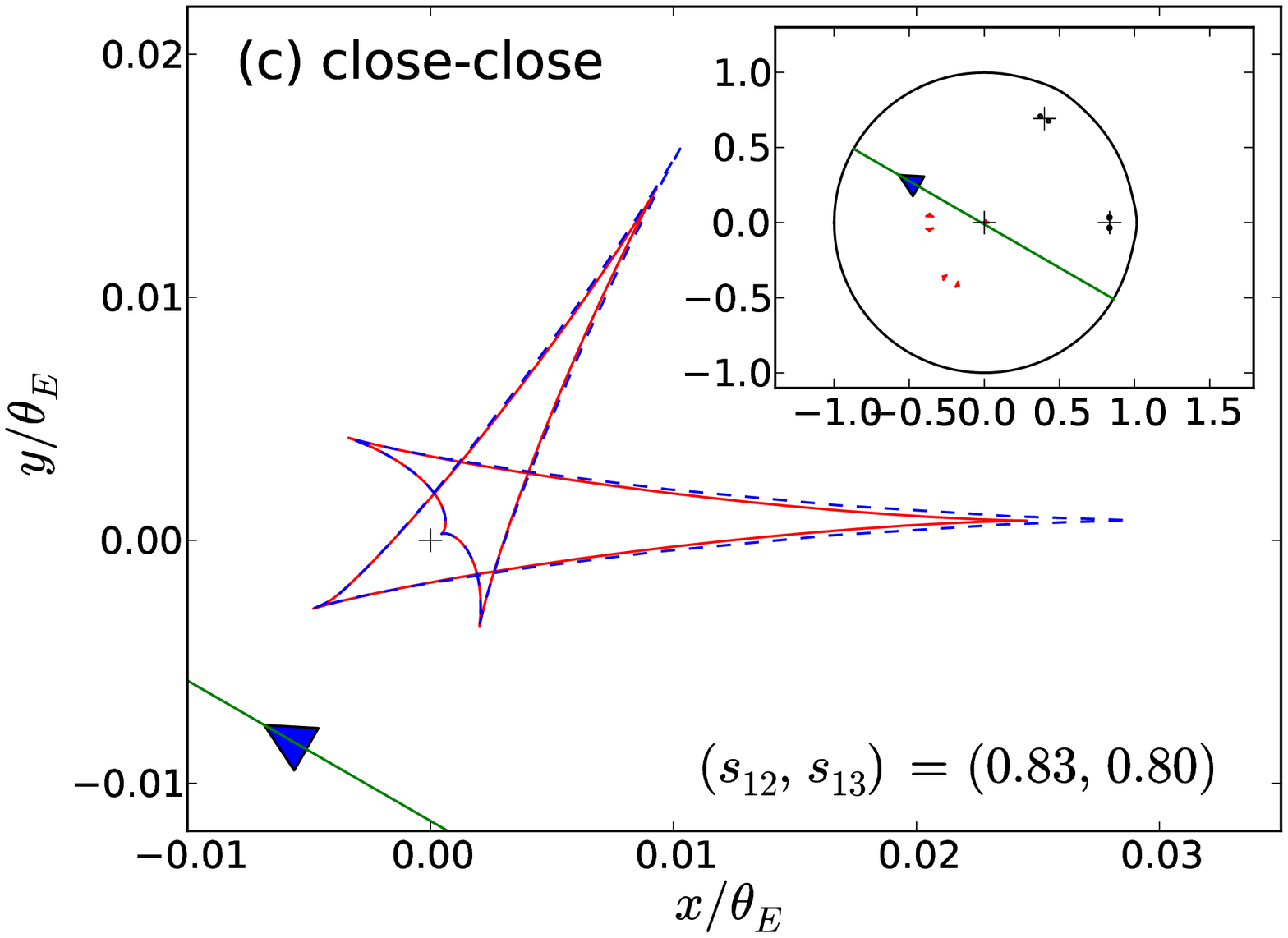}
\includegraphics[width=.45\textwidth]{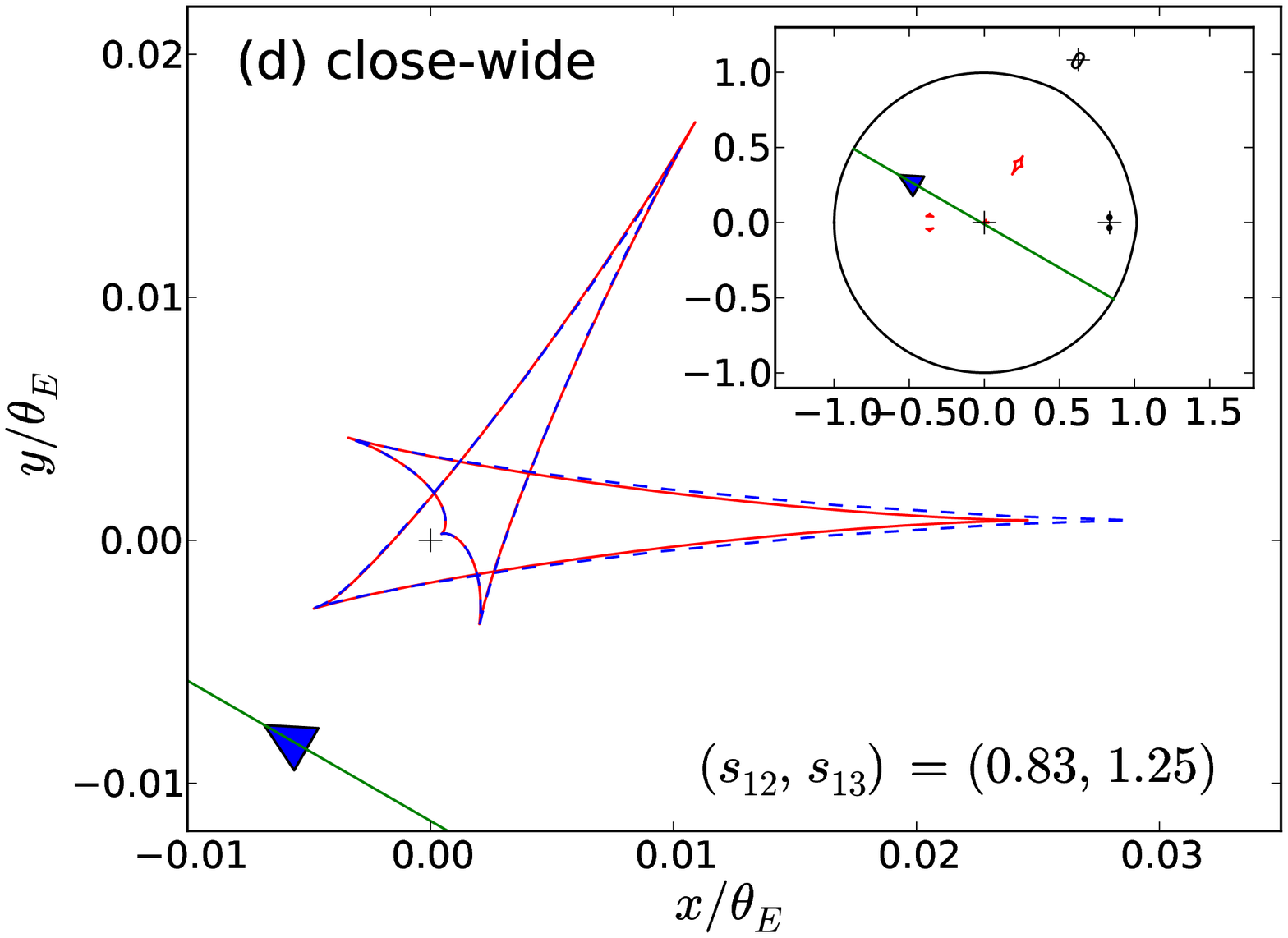}
\caption{Example of the planetary 4-fold degeneracy. Here, (i) shows the overall light curve, and (ii) zoom-in views around the peak region. The bottom parts of these figures show the residual according to the wide-wide case. (a) to (d) shows the central caustics drawn by two methods: the numerical method (red solid curves) and the linear approximation by equation~(\ref{pcu2}) (blue dashed curves). The straight green lines with an arrow are the trajectories. The static triple-lens parameters are $(q_{21}, q_{31}, \varphi)$ = (0.001, 0.001, $60^\circ$) with four cases: (a) wide-wide, $(s_{12}, s_{13}) = (1.20, 1.25)$; (b) wide-close, $(s_{12}, s_{13}) = (1.20, 0.80)$; (c) close-close, $(s_{12}, s_{13}) = (0.83, 0.80)$; (d) close-wide, $(s_{12}, s_{13}) = (0.83, 1.25)$. For the insets, the straight green lines with an arrow are the trajectories, the black rounded curves are the critical curves, the plus signs are the lenses and the colored curves are the caustics (which may be too small to see). All the parameters are shown in Table~\ref{tab:para}(1).}
\label{fig:4-fold}
\end{figure*}

For planetary lensing, i.e., when the mass ratio $q \ll 1$, \cite{Boz99} showed that the traditional perturbative method can be applied. Afterwards, \cite{An05} re-examined that the central caustics can be expanded into 
a symmetric representation to the linear approximation:
\begin{equation}
 {2 \over q}{\zeta_{\rm cc}}={\E}^{{\I}\phi} {\left[
 {1 \over { { \left( 1-{z_{\rm{p}}}{\E}^{-{\I}\phi} \right) }^2}}
 + {1 \over { {\left(1-{{\bz_{\rm{p}}}^{-1}}{\E}^{-{\I}\phi} \right) }^2} } - 1
 \right]},
\label{pcu1}
\end{equation}
where $\phi$ is the phase angle, $\zeta_{\rm cc}$ is the parametric form of the (linear approximation of the) central caustics and $z_{\rm p}$ is the planetary position in complex notation. Note that the shape of the central caustics remains the same when $z_{\rm p}$ is changed into $\bz_{\rm p}^{-1}$, which indicates the close/wide degeneracy.
The formalism can be generalised to the $n$ planet (plus primary) case:
\begin{equation} 
\label{pcu2}
 \zeta_{\rm cc} = {\E}^{{\I}\phi}\sum_{k=1}^{n}
 {{q_k} \over 2}
 {\left[
 {1 \over { {\left( 1-{z_k{\E}}^{-{\I}\phi} \right) }^2} }
 + { 1 \over { {\left( 1-{\bz_k}^{-1}{\E}^{-{\I}\phi} \right) }^2} } - 1
 \right]},
\end{equation}
which indicates a $2^n$ planetary close/wide degeneracy in microlensing. This degeneracy has already been found in observation \citep{Cho12}, and is consistent with the conclusion drawn by \cite{Boz00} (see his \S5.1), who used a different mathematical method. Here a simulated example is shown in Fig.~\ref{fig:4-fold}.

Fig.~\ref{fig:4-fold}(i) is the overall light curve of the wide-wide case with residuals below, and Fig.~\ref{fig:4-fold}(ii) is a zoom-in of the peak region. Fig.~\ref{fig:4-fold}(a)~to~\ref{fig:4-fold}(d) represent the central caustics in different cases. The red solid lines are drawn numerically, while the blue dashed lines are the linear approximation (equation~\ref{pcu2}) - they are quite similar but show subtle differences, which can be seen in the residual of the light curves. All the parameters used are shown in Table~\ref{tab:para}(1).

In Fig.~\ref{fig:4-fold}(ii), there are two deviating features in the residual between the close-close case (c) and the wide-wide case (a). By comparing the wide-close case (b) and the wide-wide case (a), we can conclude that the first feature ($t/t_E \sim$ $-0.010$ to 0.005) is due to the second planet ($q_{31}$). Similarly, the second feature ($t/t_E \sim$ 0.005 to 0.060) is due to the first planet ($q_{21}$) and more notable than the first one. As a result, the perturbations caused by individual planets are separated in this example, and this phenomena is consistent with the prediction by \cite{Rat02}. 

Since $q_{21}=q_{31}$ and $|s_{12}-1|<|s_{13}-1|$ (in the wide-wide case), it indicates that if a planet is closer to the Einstein radius of the star, the difference in its close/wide degeneracy will be more significant, i.e., the degeneracy is much easier to break. In addition, a heavier planet can also weaken the degeneracy. Note that, $(s_{12},\,s_{13})$ are chosen strictly by the $s$-$s^{-1}$ law. However, when fitting the real data, we can obtain even better degenerate solutions around these values by small changes, and thus the close/wide degeneracy may be stronger in practice.

\subsection{The external shear equations}
\label{sec:shear}

When all the other lenses are much farther away from the Einstein radius of the primary lens, the external shear approximation \citep{CR84, Dom99} is valid. Here, we rewrite the lens equation~(\ref{lensEq}) in triple-lens case ($N = 3$), and Taylor-expand the deflection terms caused by two of the masses ($m_2$ \& $m_3$) at the location of the primary mass ($m_1$)
\begin{align}  
\label{Taylor}
 \zeta = & z - {m_1 \over \bz - \bz_1} - {m_2 \over \bz - \bz_2} - {m_3 \over \bz - \bz_3}\nonumber\\
       = & z-{m_1 \over {\bz - \bz_1}} + \left( {m_2 \over {\bz_2 - \bz_1}} + {m_3 \over {\bz_3 - \bz_1}}\right) \nonumber\\
       & +\sum_{k=1}^{\infty} {\left[{m_2 \over \left(\bz_2 - \bz_1\right)^{k+1}} + {m_3 \over \left(\bz_3 - \bz_1\right)^{k+1}}\right] \left(\bz - \bz_1\right)^{k}},
\end{align}
provided $|\bz - \bz_1| \ll \min \left(|\bz_2 - \bz_1|,~|\bz_3 - \bz_1|\right)$. It is easy to transform equation~(\ref{Taylor}) into a specific form which describes a point-mass lens under perturbation
\begin{equation}  \label{perturb}
 \omega = w - {1 \over \bar{w}} 
           + \sum_{k=0}^{\infty} { {\gamma_k} \bar{w}^{k+1}},
\end{equation}
where
\begin{gather}
 \omega = {{\zeta - \zeta_1} \over \sqrt{m_1}}, \qquad w = {{z - z_1} \over \sqrt{m_1}}, \\
 \zeta_1 = z_1 + {m_2 \over {\bz_2 - \bz_1}} + {m_3 \over {\bz_3 - \bz_1}}, \label{shift} \\
 \gamma_k = \left[{m_2 \over \left(\bz_2 - \bz_1\right)^{k+2}} + {m_3 \over \left(\bz_3 - \bz_1\right)^{k+2}}\right] m_1^{k \over 2}. \label{shear}
\end{gather}
If all $\gamma_k$ are equal to each other for two lens systems, it will be a perfect degeneracy. However, this would also imply two systems are identical. Instead of the perfect but trivial degeneracy, we can identify approximate degeneracies by truncating equation~(\ref{perturb}) after $k=1$ (see \citealt{An05})
\begin{equation}
\label{gammaEq0}
 |\gamma_0'| = |\gamma_0|, \qquad {\gamma_1' \over \gamma_0'} = {\gamma_1 \over \gamma_0 },
\end{equation}
and such degeneracies may be indistinguishable within observational uncertainty. Throughout this paper, we use primed symbols for the parameters of derived degenerate systems, to differentiate them from those of the initial system. Note that in this case, three component equations in equations~(\ref{gammaEq0}) are responding to five lens parameters named $(m_{21},\,m_{31},\,s_{12},\,s_{13},\,\varphi)$, and one relative parameter is needed when comparing two different lens systems. 

So it is convenient to include another parameter to determine the relative orientation between each triple-lens system, and we find the direction angle $\theta$ is quite suitable. By choosing parameter set $(m_{21},\,m_{31},\,s_{12},\,s_{13},\,\varphi,\,\theta)$, we can always make the first two $\gamma_k$ terms to be the same for two different lens systems, i.e., 
\begin{equation}
\label{gammaEq}
 \gamma_0' = \gamma_0  \equiv c_1, \qquad \gamma_1' = \gamma_1 \equiv c_2,
\end{equation}
where $c_1$ and $c_2$ are two complex constants. Now, there are four equations to determine six free parameters in a triple-lens system, implying a continuous degeneracy with two remaining parameters $\varphi$ and $\theta$. To simplify calculation, we also choose the position of the primary lens mass ($m_1$) as the origin of the coordinate system. Hence, $z_1 = 0$, $z_2 = s_{12}\E^{\I\theta}$ and $z_3 = s_{13}\E^{\I(\varphi+\theta)}$. As a result, the external shear equations for triple-lens can be derived from equations~(\ref{shear}) and~(\ref{gammaEq}) as 
\begin{subequations} 
\label{3shearEq}
\begin{align}
 {m_2 \over s_{12}^2} \cos{2\theta} + {m_3 \over s_{13}^2} \cos{2\left(\varphi+\theta\right)} & \equiv a_1,  \\
 {m_2 \over s_{12}^2} \sin{2\theta} + {m_3 \over s_{13}^2} \sin{2\left(\varphi+\theta\right)} & \equiv b_1, \label{3shearEq:b} \\
 {m_2 \over s_{12}^3} \cos{3\theta} + {m_3 \over s_{13}^3} \cos{3\left(\varphi+\theta\right)} & \equiv {a_2 \over \sqrt{m_1}}, \\
 {m_2 \over s_{12}^3} \sin{3\theta} + {m_3 \over s_{13}^3} \sin{3\left(\varphi+\theta\right)} & \equiv {b_2 \over \sqrt{m_1}}, \label{3shearEq:d}
\end{align}
\end{subequations}
where $c_1 = a_1 + {\I}b_1$, $c_2 = a_2 + {\I}b_2$. Note that both the initial and derived triple-lens parameters satisfy these equations. 

In principle, there are two steps to obtain all potential continuous degenerate solutions. Firstly, $c_1$ and $c_2$ are calculated by equations~(\ref{3shearEq}) with the initial parameters $(m_2,\,m_3,\,s_{12},\,s_{13},\,\varphi,\,\theta)$. And then, equations~(\ref{3shearEq}) are called again to calculate all sets of derived parameters $(m_2',\,m_3',\,s_{12}',\,s_{13}',\,\varphi',\,\theta')$. We find that when $(\theta,\,\varphi',\,\theta')$ are chosen, equations~(\ref{3shearEq}) can be solved analytically (see Appendix~{\ref{subsec:3-3}} for the detailed procedure). Moreover, in the coordinate system determined by $\theta$, the trajectory angle should be 
\begin{equation}
 \alpha_{\theta} = \alpha_0+\theta,
\end{equation}
which is crucial when generating the degenerate light curves.

\subsection{The continuous external shear degeneracy}
\label{sec:triple}

\begin{figure*}
\graphicspath{}
\centering
\vspace{-3mm}
\includegraphics[width=.45\textwidth]{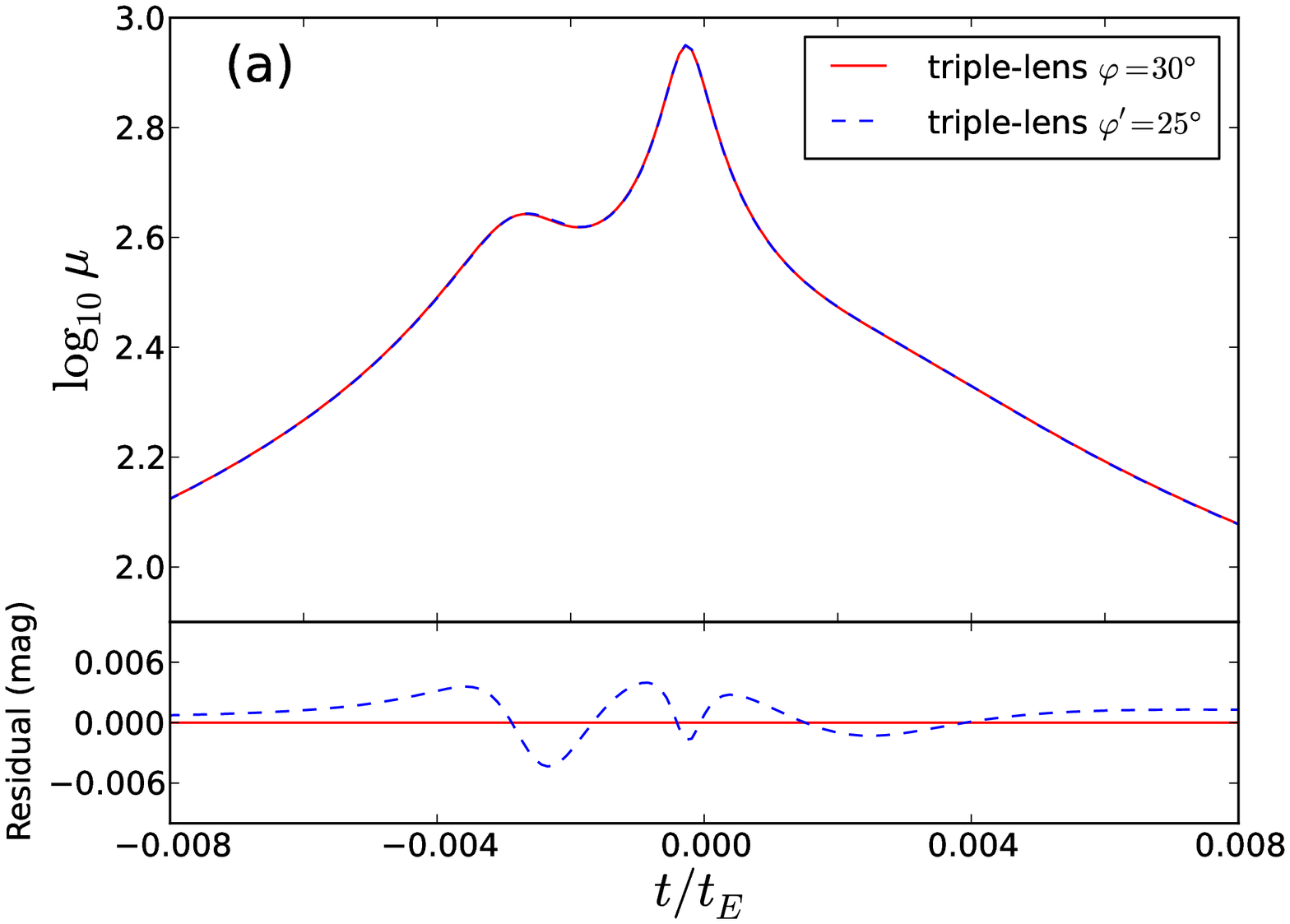}
\includegraphics[width=.45\textwidth]{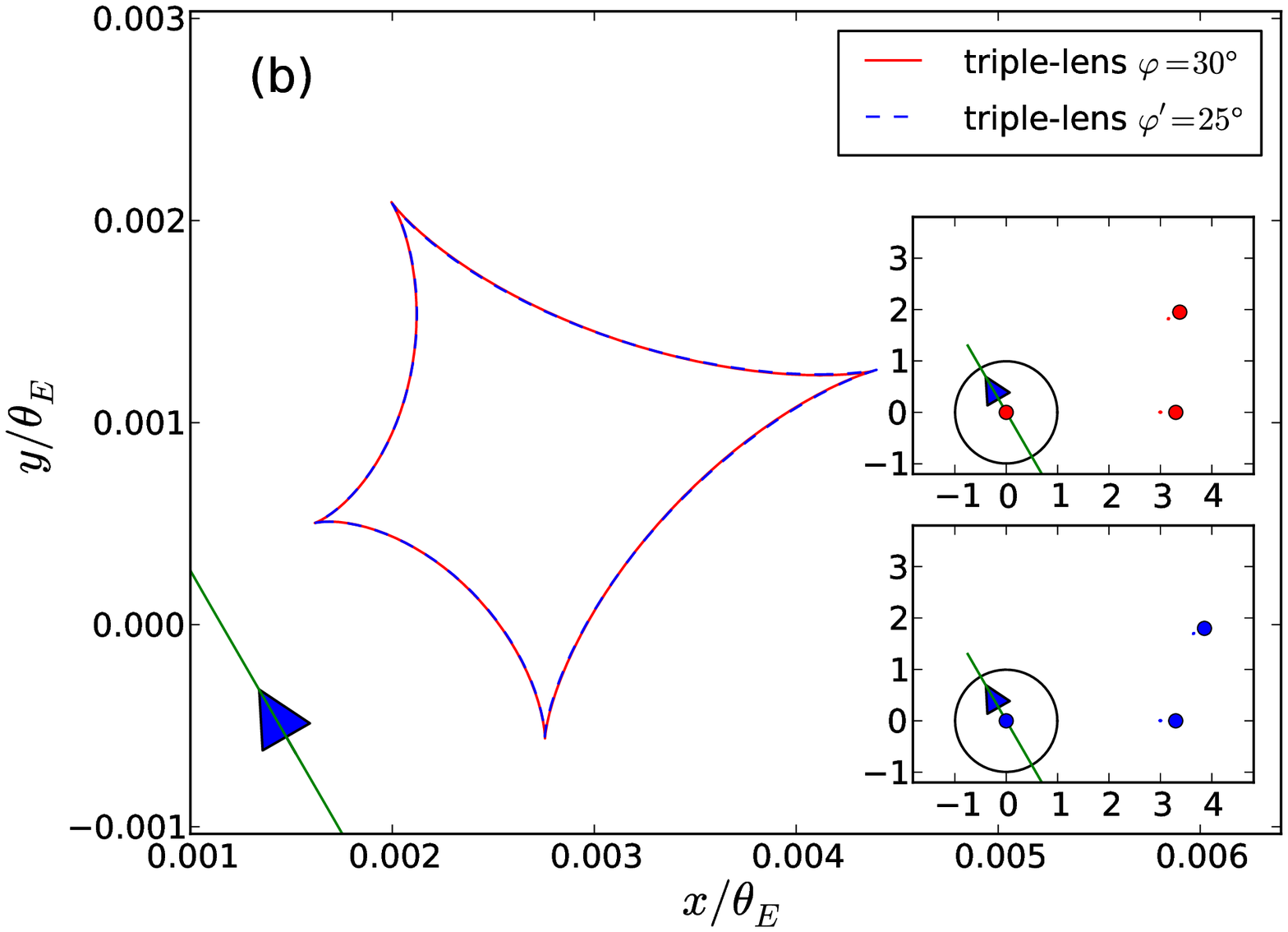}
\includegraphics[width=.45\textwidth]{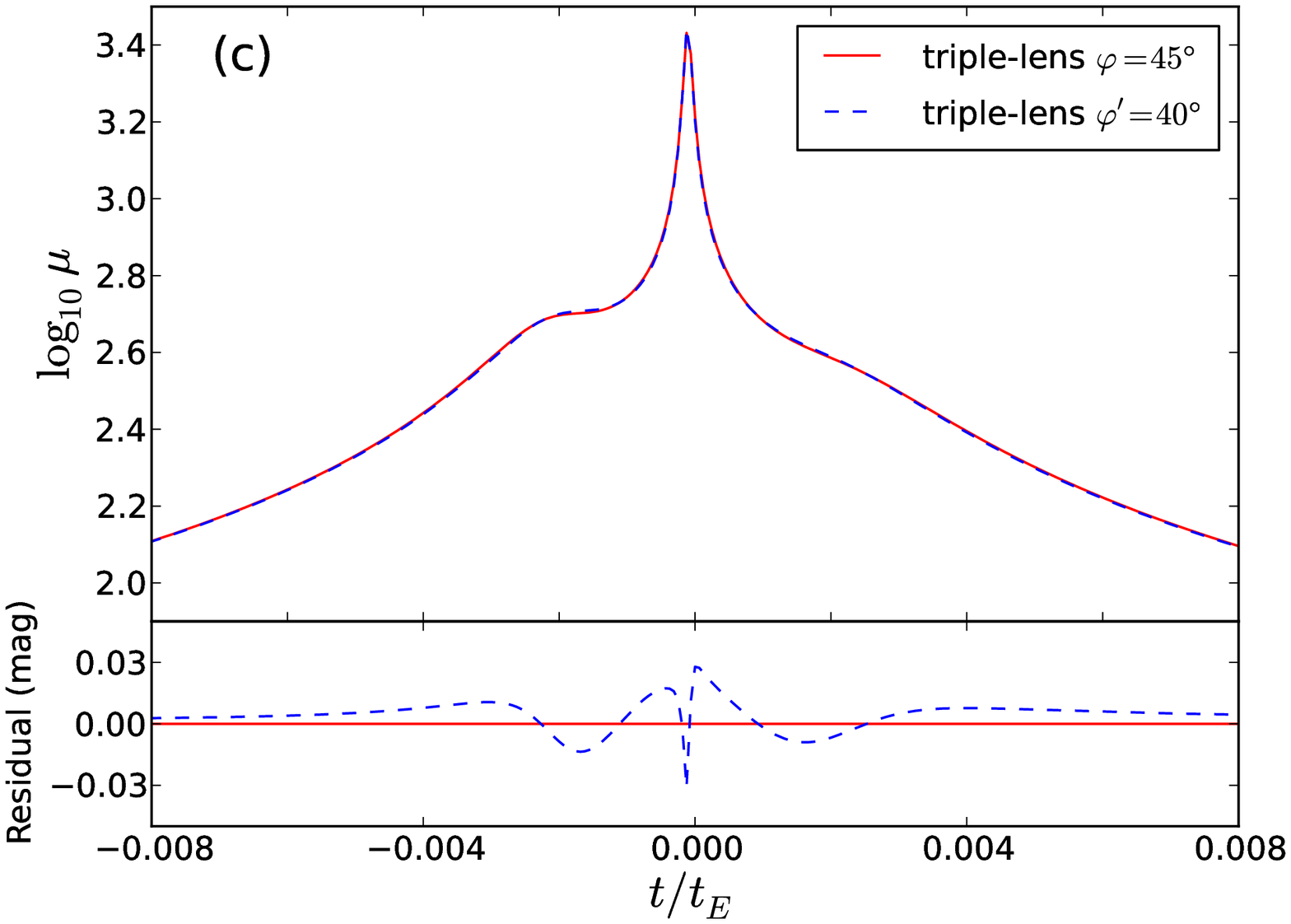}
\includegraphics[width=.45\textwidth]{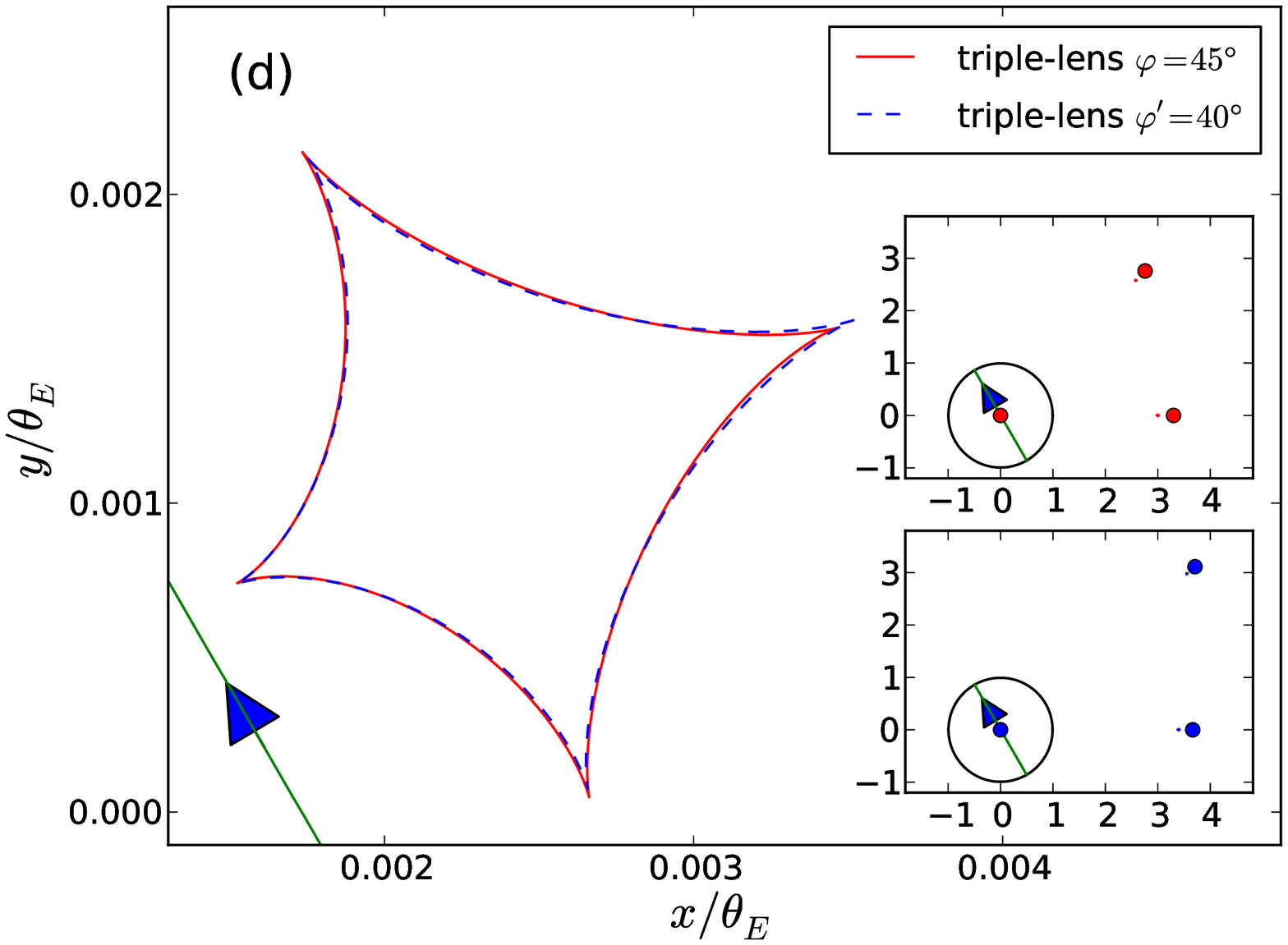}
\includegraphics[width=.45\textwidth]{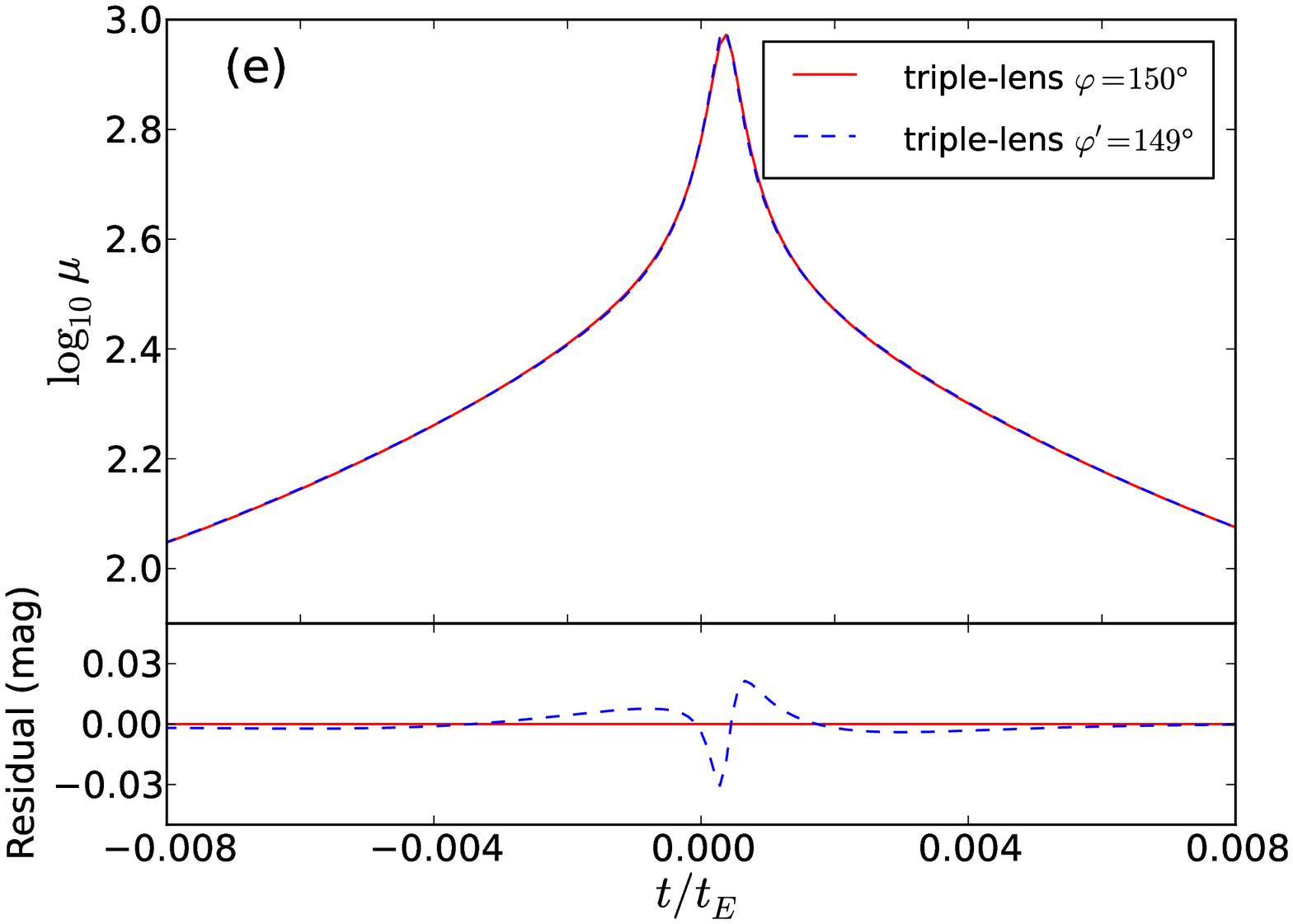}
\includegraphics[width=.45\textwidth]{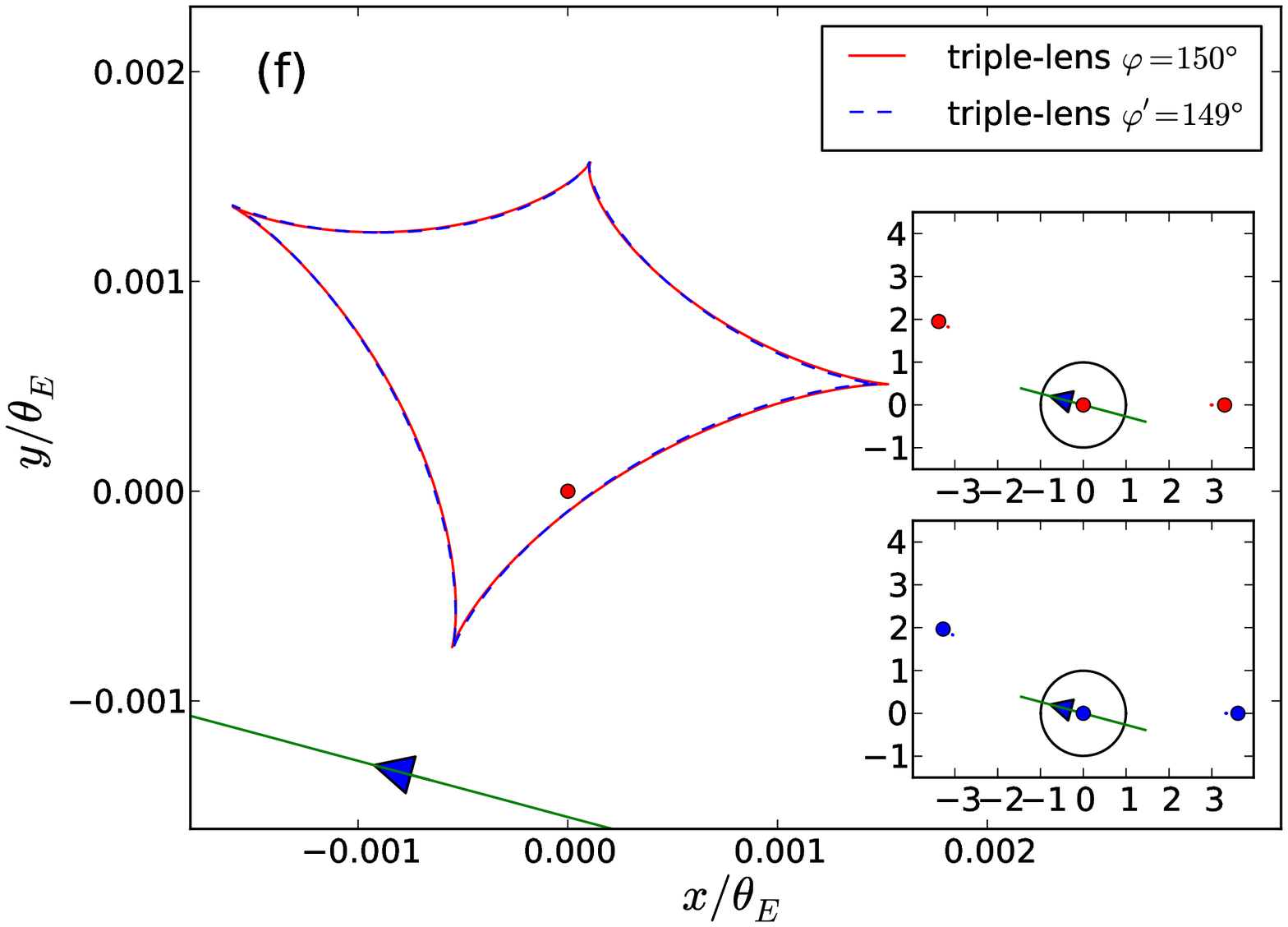}
\caption{Examples of the continuous external shear degeneracy. The left panel shows peak-region light curves with the residual between them at the bottom, while the right panel shows the central caustics with the overall configurations in the insets. The red solid lines represent the initial static triple-lens system, and the blue dashed lines are for the derived system. For the insets, the straight green lines with an arrow are the trajectories, the black rounded curves are the critical curves, the colored dots are the lenses and the colored curves are the caustics (which may be too small to see). Note that in each example, we set $\theta=\theta'=0^\circ$. All the parameters are shown in Table~\ref{tab:para}(2).}
\label{fig:triple}
\end{figure*}

\begin{figure*}
\graphicspath{}
\centering
\vspace{-3mm}
\includegraphics[width=.33\textwidth]{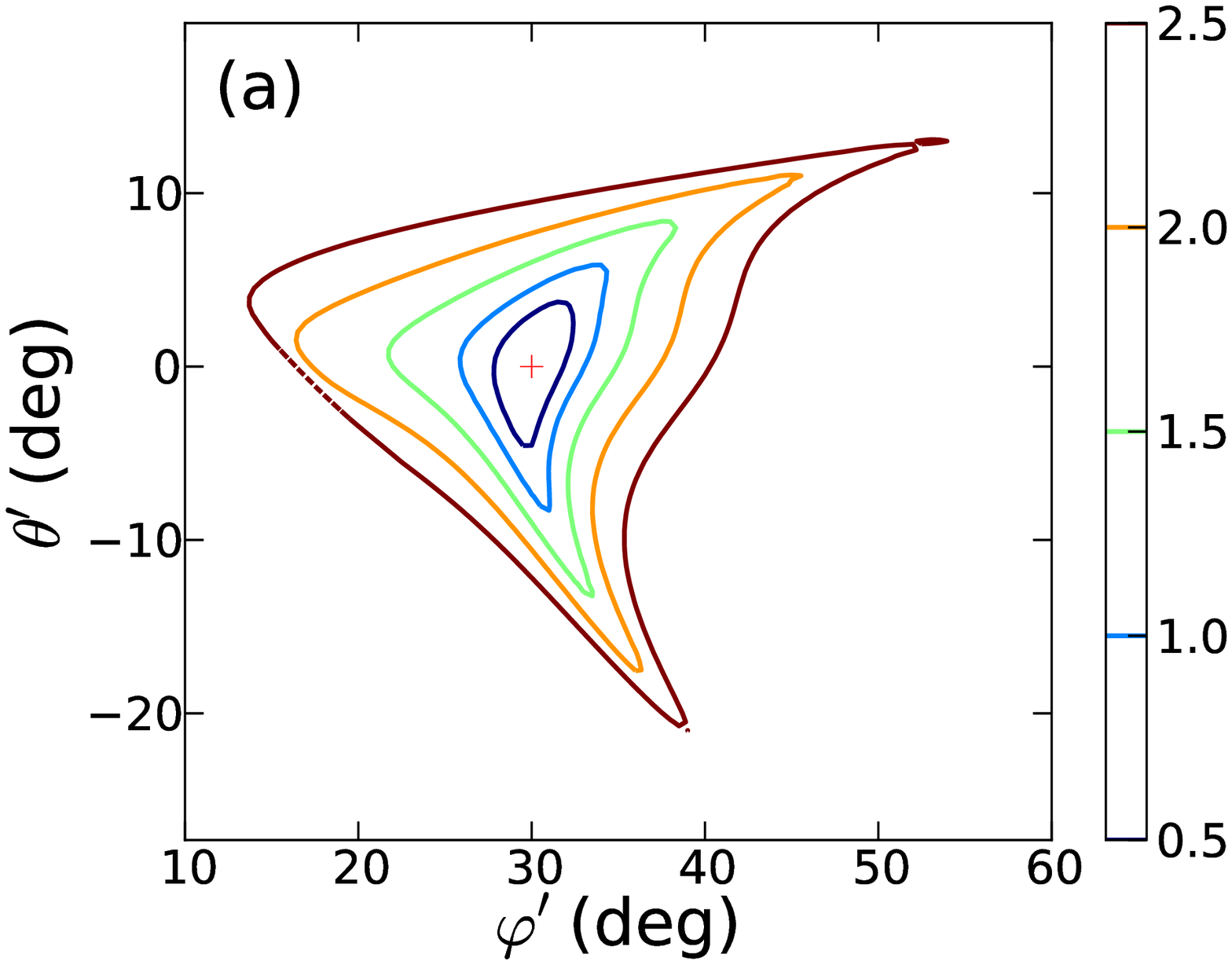}
\includegraphics[width=.33\textwidth]{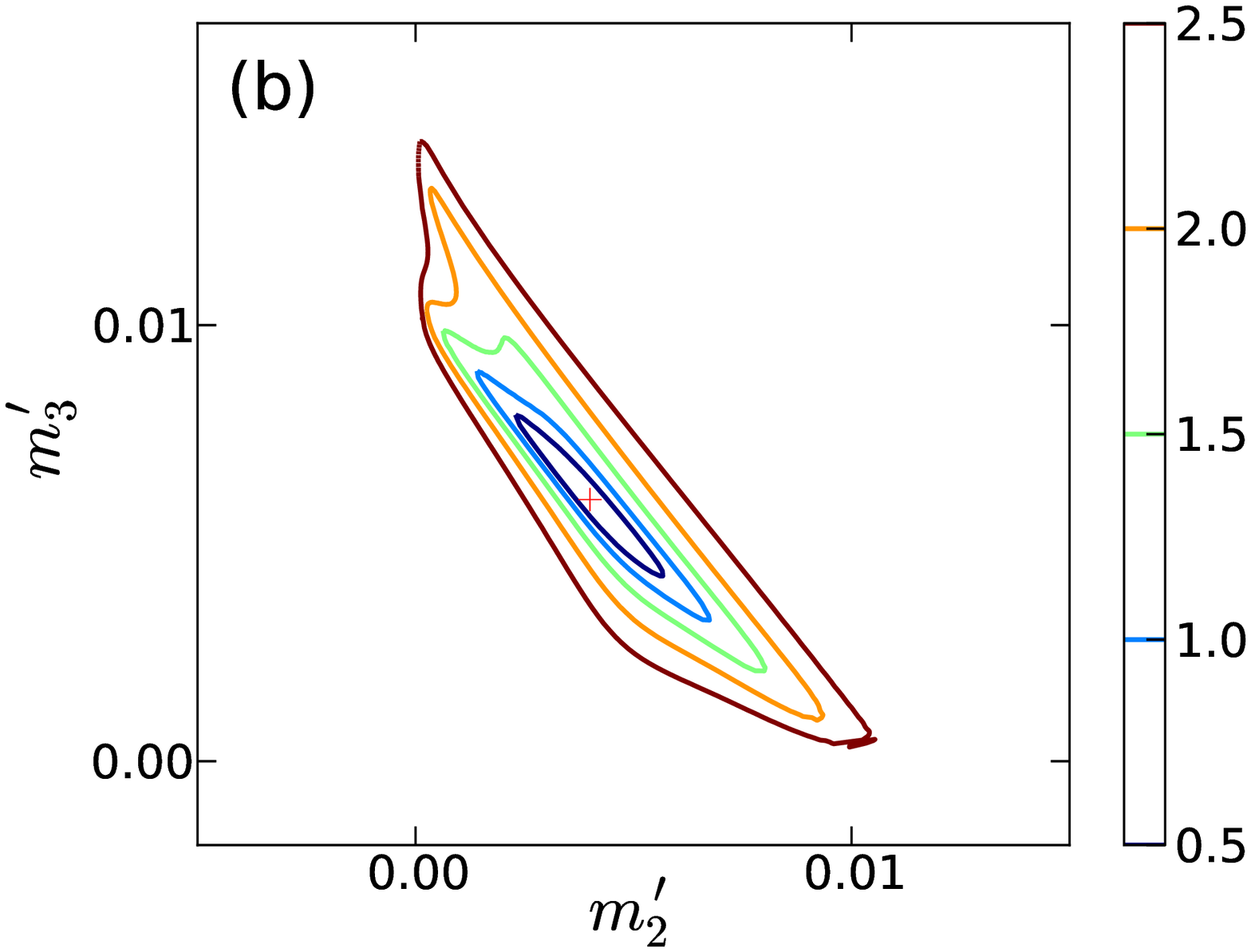}
\includegraphics[width=.33\textwidth]{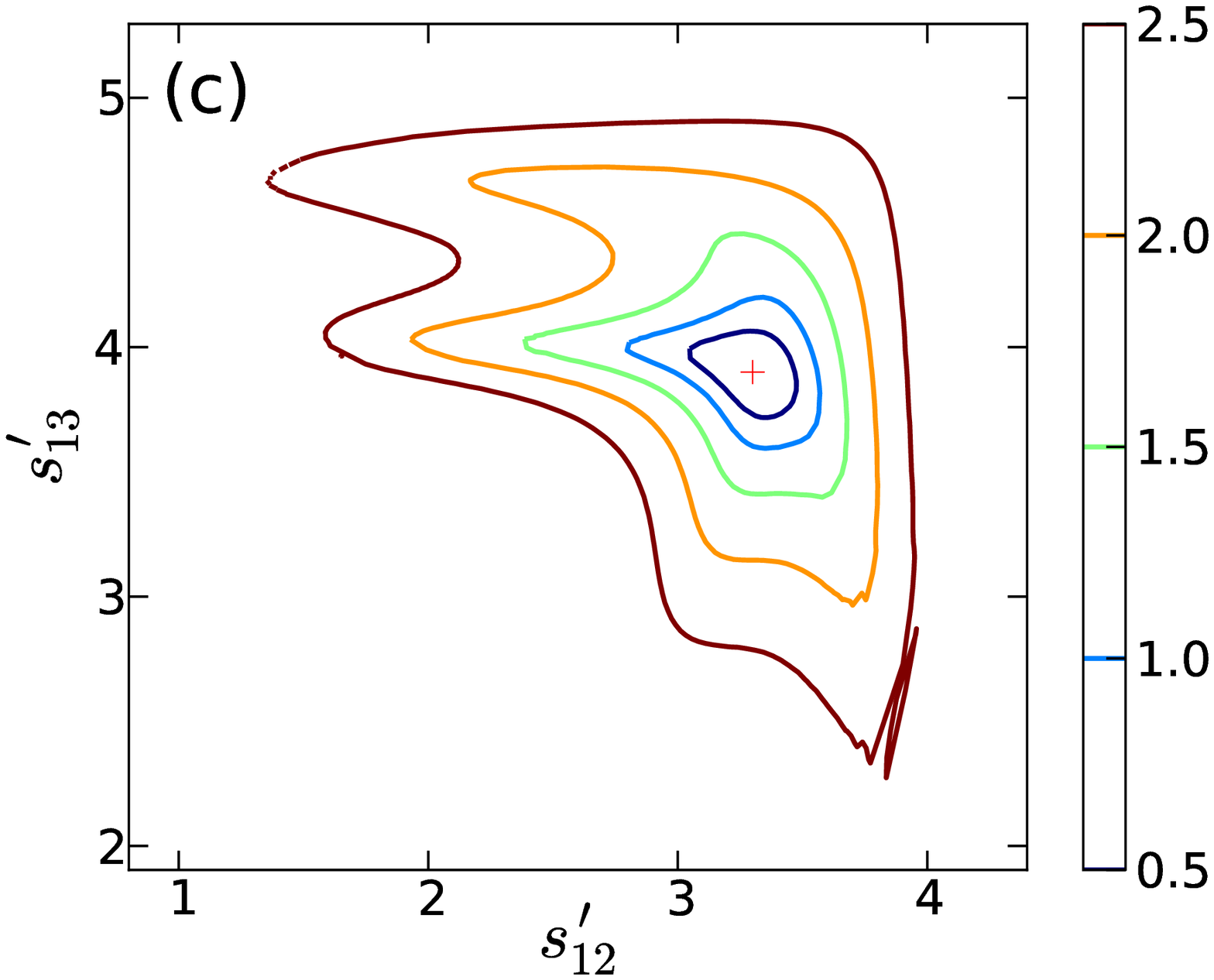}
\includegraphics[width=.33\textwidth]{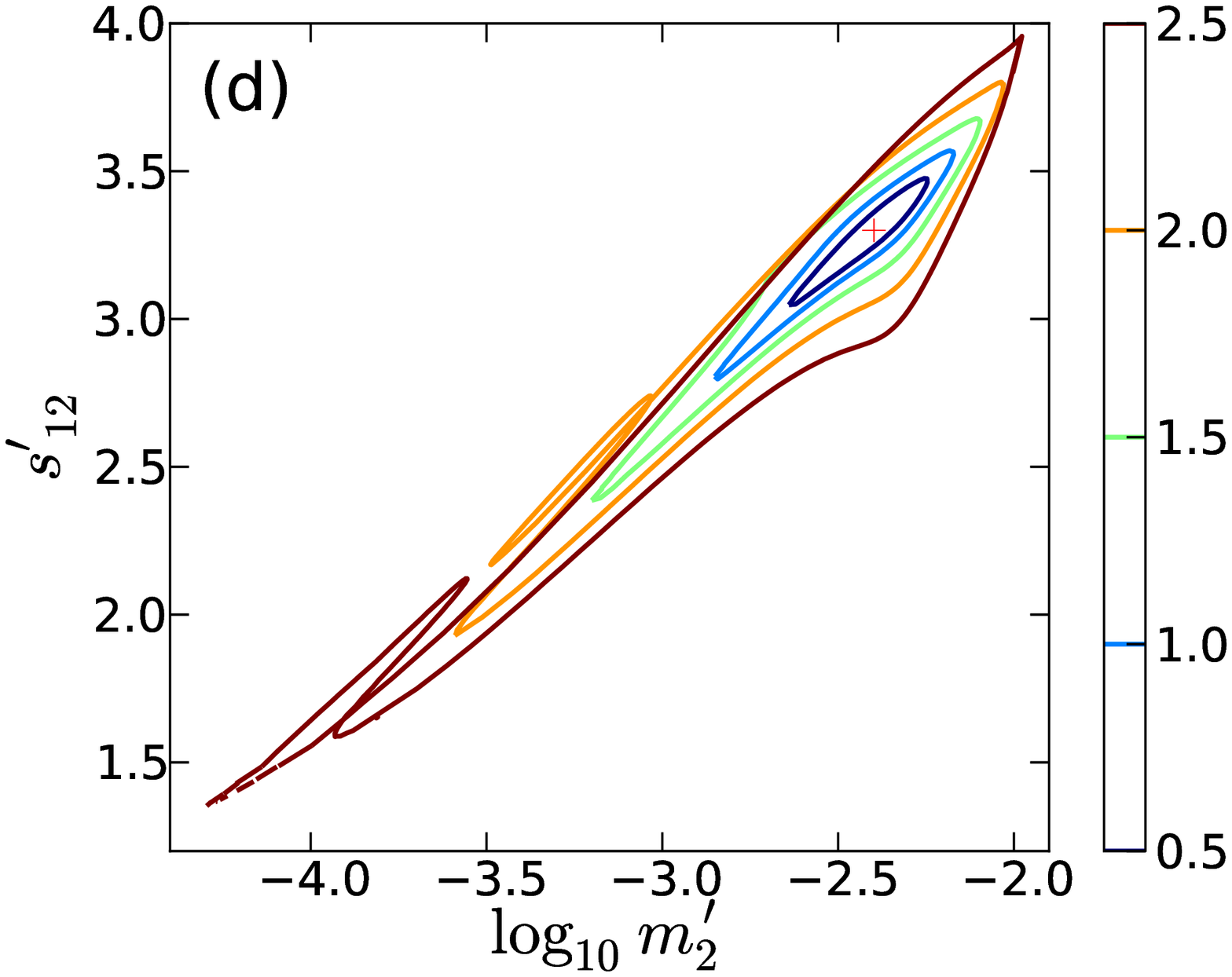}
\includegraphics[width=.33\textwidth]{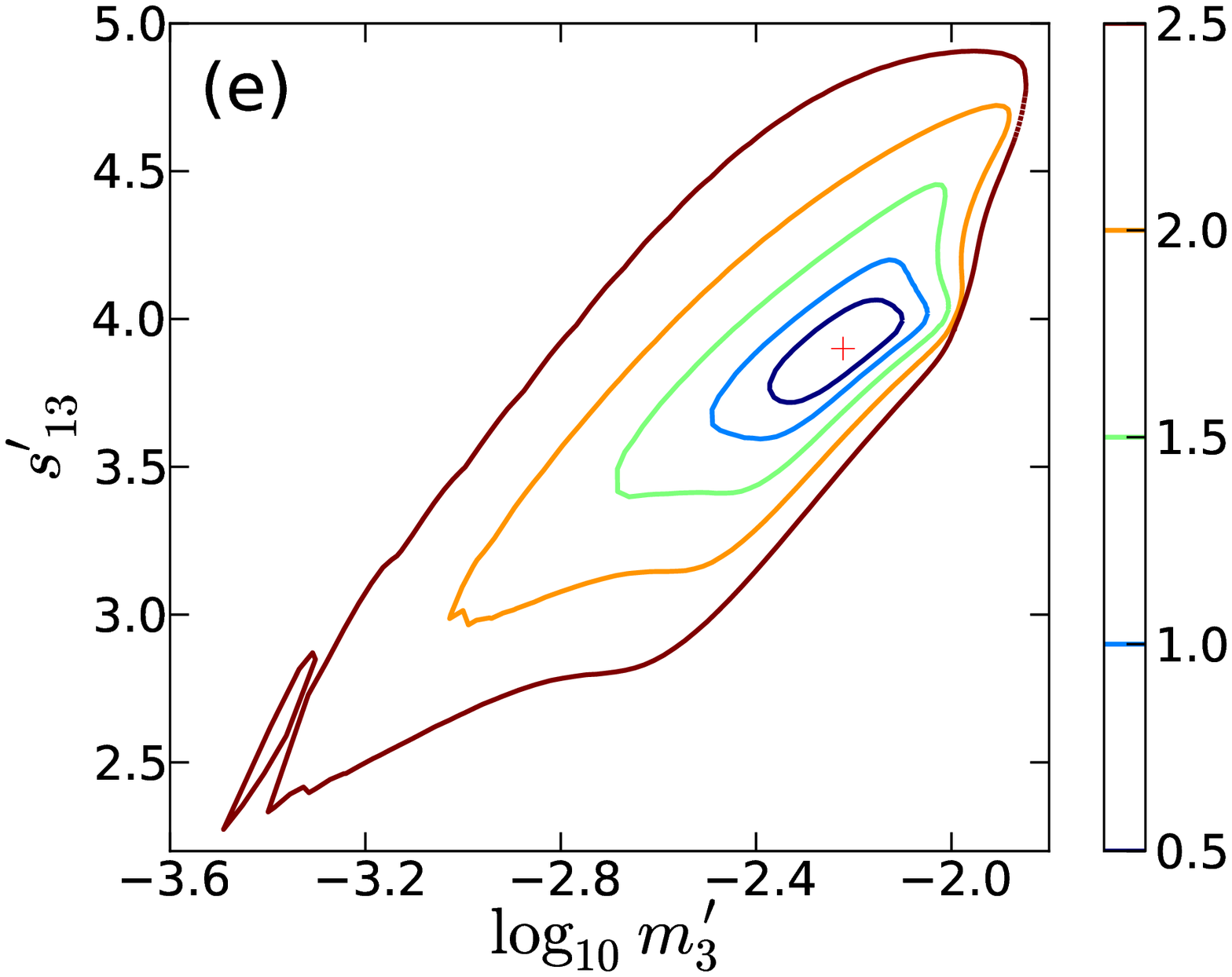}
\caption{Contours of $\Delta \chi^2$ between the input model (indicated by a cross) and degenerate models. 
The input model is shown in Fig.~\ref{fig:triple}(a) ($\varphi = 30^\circ$, $\theta = 0^\circ$). Different panels are for different combinations of lens parameters. The plus sign in each panel marks the position of the initial parameters, so it also locates $\Delta \chi^2=0$. The contour levels shown are $\log_{10}{\Delta\chi^2}=0.5,\,1.0,\,1.5,\,2.0,\,2.5$ respectively.}
\label{fig:c30}
\end{figure*}

\begin{figure*}
\graphicspath{}
\centering
\vspace{-3mm}
\includegraphics[width=.33\textwidth]{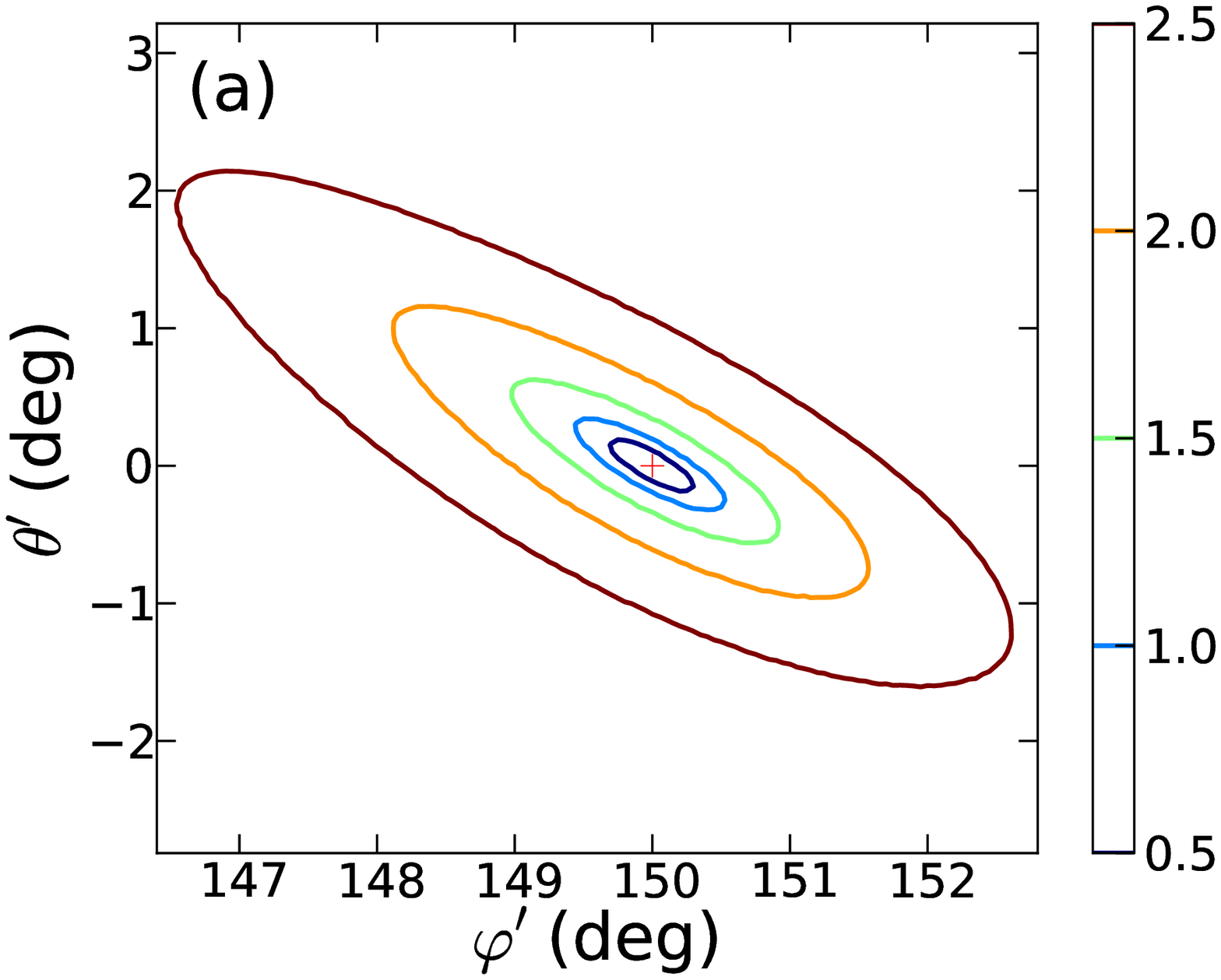}
\includegraphics[width=.33\textwidth]{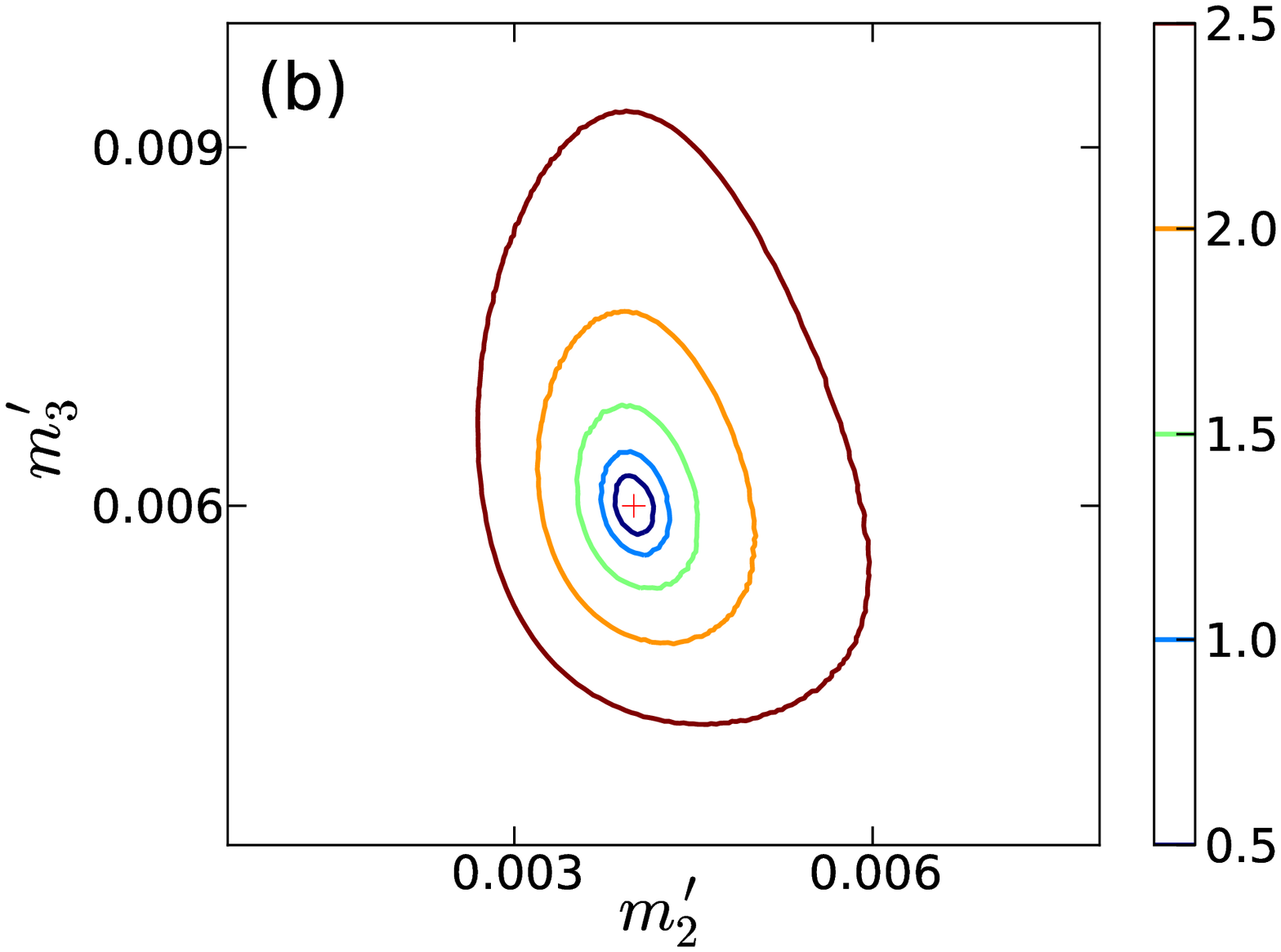}
\includegraphics[width=.33\textwidth]{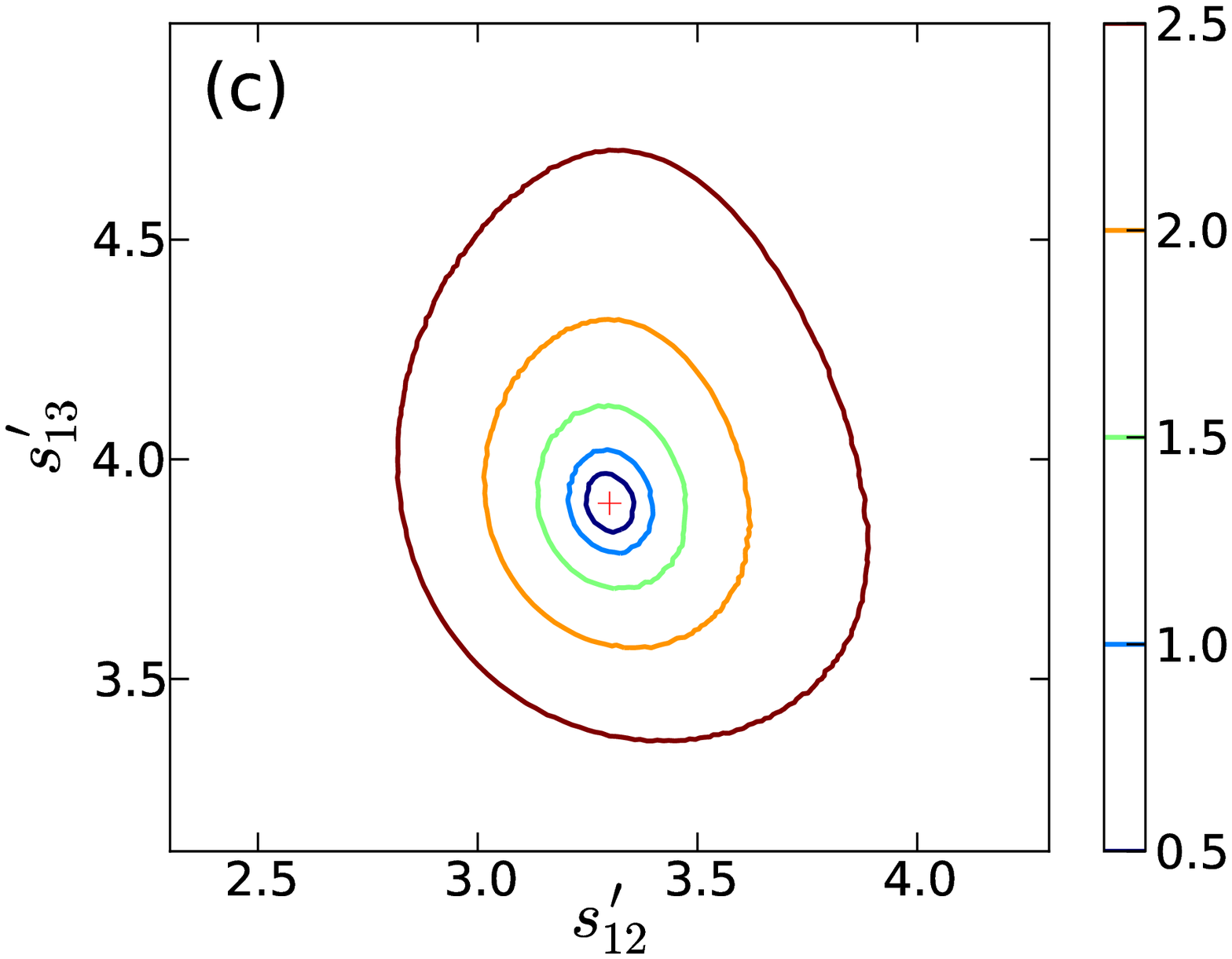}
\includegraphics[width=.33\textwidth]{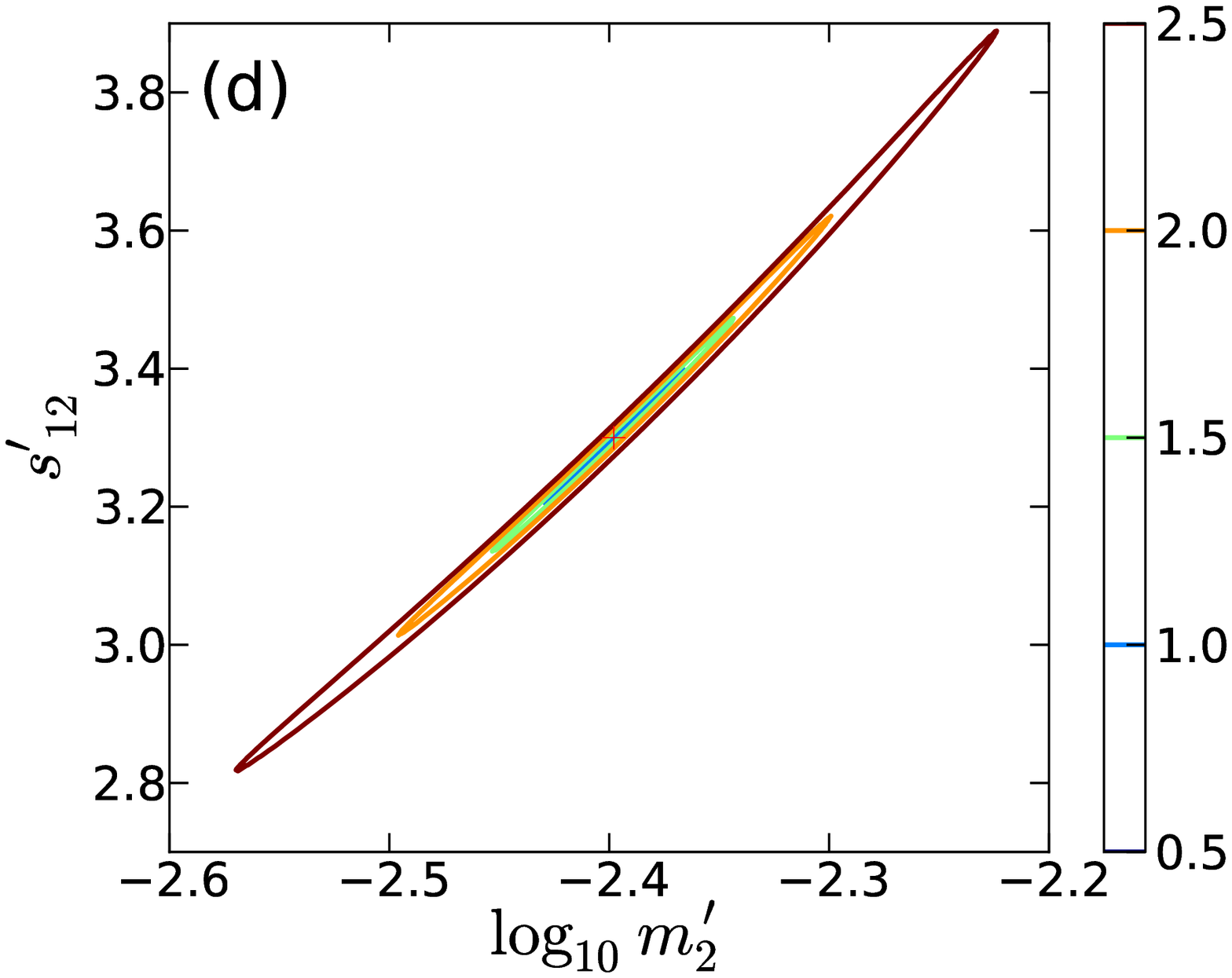}
\includegraphics[width=.33\textwidth]{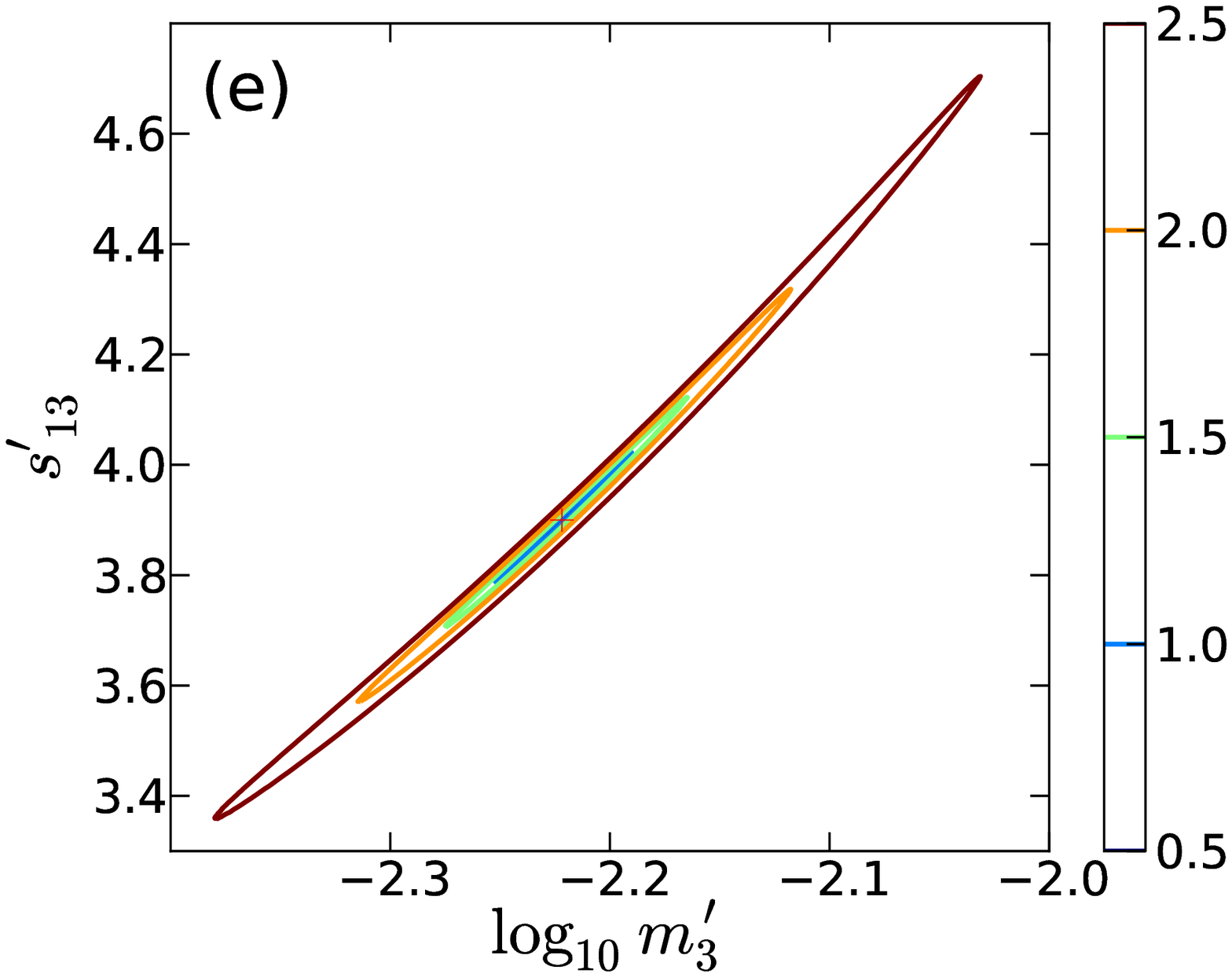}
\caption{Contours of $\Delta \chi^2$ between the input model (indicated by a cross) and degenerate models. The input model is shown in Fig.~\ref{fig:triple}(e) ($\varphi = 150^\circ$, $\theta = 0^\circ$). The symbols are the same as in
Fig. \ref{fig:c30}. The constraints are much tighter compared with those in Fig. \ref{fig:c30} (see \S\ref{sec:triple}).
}
\label{fig:c150}
\end{figure*}

To check the reliability of the truncation in equation~(\ref{perturb}), higher-order effects should be considered, i.e., for $k \geq 2$, if we have
\begin{equation}
 \left|{\gamma_k' \over \gamma_1'} - {\gamma_k \over \gamma_1}\right| \ll 1,
\end{equation}
then there should be a set of ``continuous degeneracies'' for different triple-lens systems. 

For the assumptions and approximation discussed in \S\ref{sec:shear}, equation~(\ref{perturb}) is suitable to describe the shape of the central caustics. Although they turn out to be much smaller than the ones near other lower mass objects (i.e., planetary caustics), these caustics play dominant roles in high magnification events, which are particularly important in the current mode of discovering exoplanets where a combination of surveys and followups is used (\citealt{G&S98}). 

Fig.~\ref{fig:triple} shows three examples of the external shear continuous degeneracies in static triple-lens case. The red solid lines are for the initial system, while the blue dashed lines represent one of the continuous degenerate systems whose parameters are calculated by equation~(\ref{3shearEq}). The left panel shows the comparison around the peak region, and the right panel shows the comparison of the central caustics. Note that we have already shifted the blue dashed caustic as suggested by equation~(\ref{shift}) to overlap those two caustics together, and set $\theta'= \theta=0^\circ$ in the figures for convenience. All the parameters are shown in Table~\ref{tab:para}(2). As shown in these examples, the external shear degeneracies exist in triple-lens case, although they may not always have the same strength. 

In the simulated events, the magnifications are very high $\mu \sim 10^3$, still when the angles are not so different (see the top panel), the differences are only $\sim 0.005$ mag lasting for $\sim 0.005 t_E$ (approximately a few hours for a typical event), which may be difficult to detect even using the next generation microlensing event. For the other two panels, the differences are somewhat larger, reaching $0.03$ mag and $0.05$ mag respectively.

We have simulated many more events, and find some trends between the degenerate strength and the input parameters  $(\varphi,\,\theta,\,\Delta\varphi,\,\Delta\theta)$, where $\Delta\theta = \theta' - \theta$ and $\Delta\varphi = \varphi' - \varphi$. The correlation between the degenerate strength  and these parameters are as follows:
\begin{enumerate}
\item When $\theta = 0^\circ$, there are some  $\varphi$ values which may make the analytic method invalid. First of all, when $\varphi = 0^\circ$ or $180^\circ$, equations~(\ref{3shearEq:b}) and (\ref{3shearEq:d}) are equal to 0. To get a non-trivial solution, the only choice is to set $m_3' = 0$, which means that the initial triple-lens system will be degenerate with a double-lens system. We will discuss this in more detail in \S\ref{sec:double}. Secondly, when $\varphi = 90^\circ$, equation~(\ref{3shearEq:b}) is equal to 0. We have either $m_3' = 0$ or $\sin{2\varphi'} = 0$, leading to a trivial solution. Similarly, when $\varphi = 60^\circ$ or $120^\circ$, equation~(\ref{3shearEq:d}) is equal to 0. As a result, the degeneracy will vanish around $\varphi = 60^\circ$, $90^\circ$ and $120^\circ$.
\item
 In fact, these discreet $\varphi$ values divide the whole characteristic angular space into four regions, i.e., $(0^\circ,\,60^\circ)$, $(60^\circ,\,90^\circ)$, $(90^\circ,\,120^\circ)$ and $(120^\circ,\,180^\circ)$. The degeneracy strength behaves differently in these regions. For $\varphi \in (0^\circ,\,60^\circ)$, a smaller $\varphi$ makes the degeneracy stronger, which can be seen by comparing the first two examples in Fig.~\ref{fig:triple}. Both having $\Delta\varphi = -5^\circ$ and $\Delta\theta = 0^\circ$, the degeneracy of the first example is stronger because a smaller $\varphi$ ($30^\circ < 45^\circ$). However, for $\varphi \in (120^\circ,\,180^\circ)$, this correlation is weak and even reversed when $\varphi$ approaches $180^\circ$. Finally, for $(90^\circ,\,120^\circ)$ and $(120^\circ,\,180^\circ)$, the degeneracy is normally weak although becomes stronger when $\varphi$ is $75^\circ$ or $105^\circ$.
\item
In addition, an acute $\varphi$ has a stronger degeneracy than its complementary ($180^\circ-\varphi$), i.e., there is no symmetry about $\varphi = 90^\circ$. For example, with $\varphi = 30^\circ$ in the first example (Fig.~\ref{fig:triple}a~and~\ref{fig:triple}b), the residual between the light curves is quite small even though $\Delta\varphi = -5^\circ$. But when $\varphi = 150^\circ$ (Fig.~\ref{fig:triple}e~and~\ref{fig:triple}f), $\Delta\varphi = -1^\circ$ leads to far more significant deviations.
\item
Not surprisingly, the degeneracy becomes stronger if $\Delta\varphi \rightarrow 0^\circ$ and $\Delta\theta \rightarrow 0^\circ$. Besides, there seems no symmetry about $\Delta\varphi = 0^\circ$: for $\varphi \in (0^\circ,\,60^\circ)$ and $\varphi \in (120^\circ,\,180^\circ)$, the degeneracies with $\Delta\varphi < 0^\circ$ are always better than those with $\Delta\varphi > 0^\circ$; while for $\varphi \in (60^\circ,\,90^\circ)$ and $\varphi \in (90^\circ,\,120^\circ)$, the opposite is true. Actually, all the examples in Fig.~\ref{fig:triple} are simulated with $\Delta\varphi < 0^\circ$ and $\Delta\theta = 0^\circ$.
\end{enumerate}

To illustrate how different parameters may be correlated in a real event, we simulate a light curve
covering a duration of $\Delta t = 3.0\,t_{\rm E}$ with 8642 data points (corresponding to a cadence of 10 min for a typical microlensing event with $t_{\rm E}=20$ d). The $\chi^2$ is given by
\begin{equation}
 \chi^2 = \sum{\left(m_i-m_{oi}\right)^2 \over \sigma_{oi}^2}, 
\end{equation}
where $m_{oi}$ and $m_i$ are calculated by the initial and degenerate parameters respectively, and $\sigma_{oi}$ is taken to be
\begin{equation}
 \sigma_{oi}^2 = {0.05^2 \over \mu_{oi}} + 0.003^2, 
\end{equation}
where 0.05 is the baseline magnitude error, and 0.003 is the assumed systematic error. The scaling with $\mu_{oi}$ takes into account of the Poisson statistics due to magnification. The error model here is somewhat realistic, but should be taken as illustrative.

Figs.~\ref{fig:c30} and~\ref{fig:c150} show the correlations between the derived parameters of certain triple-lens systems for $(\varphi, \theta)=(30^\circ, 0^\circ)$ and $(150^\circ, 0^\circ)$ respectively. We show the $\chi^2$ contours between the input model and other models with $\log_{10}{\Delta\chi^2}=0.5,\,1.0,\,1.5,\,2.0,\,2.5$. The $\chi^2$ contours have some interesting features. For $(\varphi, \theta)=(30^\circ, 0^\circ)$ , the $\theta'$ and $\varphi'$ parameters appear to show triangle shapes, while the $m_2^\prime$ and $m_3^\prime$ follow roughly a straight line with the same total mass. The $m_2'$ - $s_{12}'$ and $m_3'$ - $s_{13}'$ follow roughly lines with constant $m_2'/s_{12}'^2$ and $m_3'/s_{12}'^2$. For $(\varphi, \theta)=(150^\circ, 0^\circ)$, the contours are much tighter, as we discussed above, but the trends remain roughly the same as the case of $(\varphi, \theta)=(30^\circ, 0^\circ)$.

\subsection{The double-triple lens degeneracy}
\label{sec:double}

\begin{figure*}
\graphicspath{}
\centering
\vspace{-3mm}
\includegraphics[width=.45\textwidth]{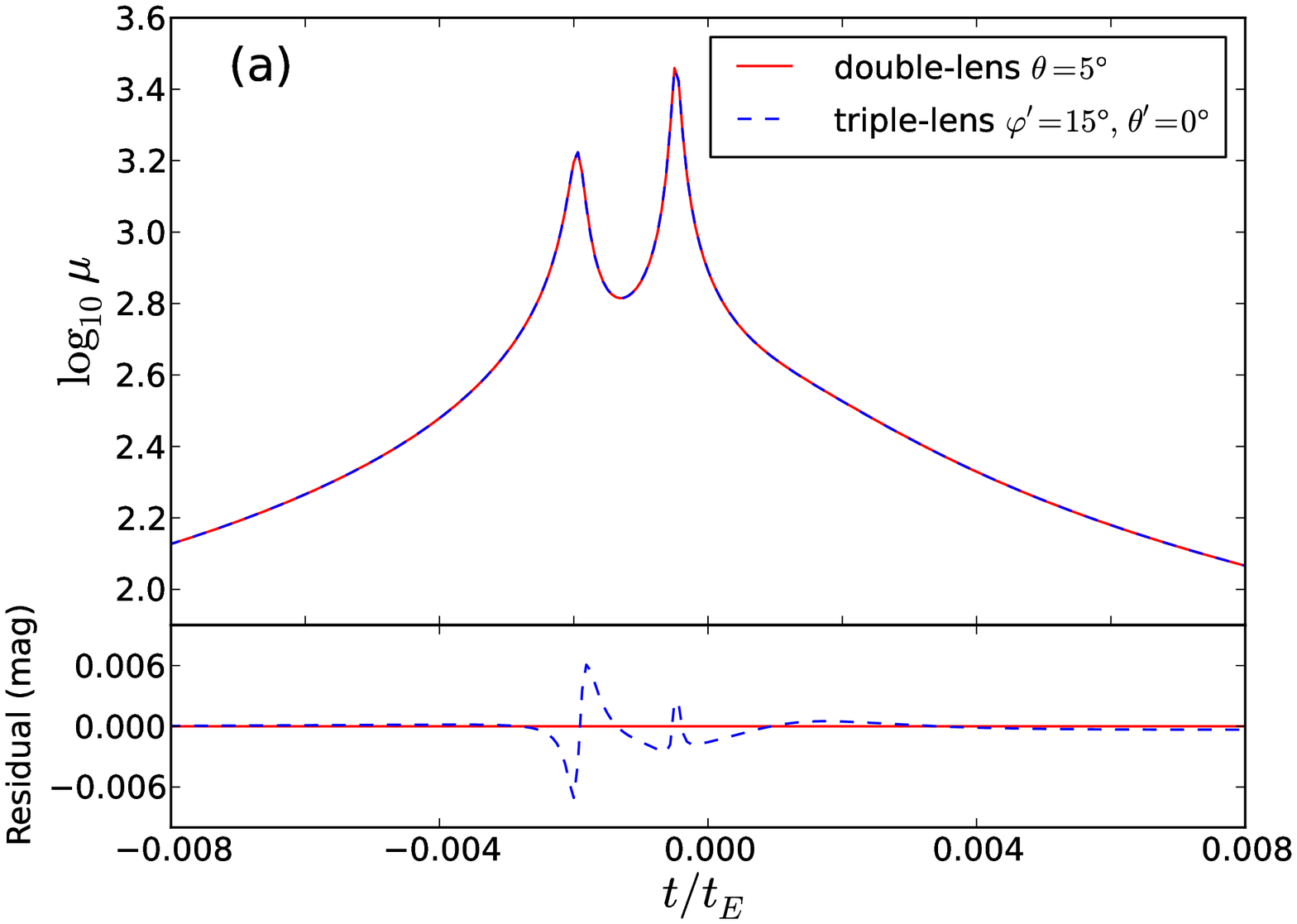}
\includegraphics[width=.45\textwidth]{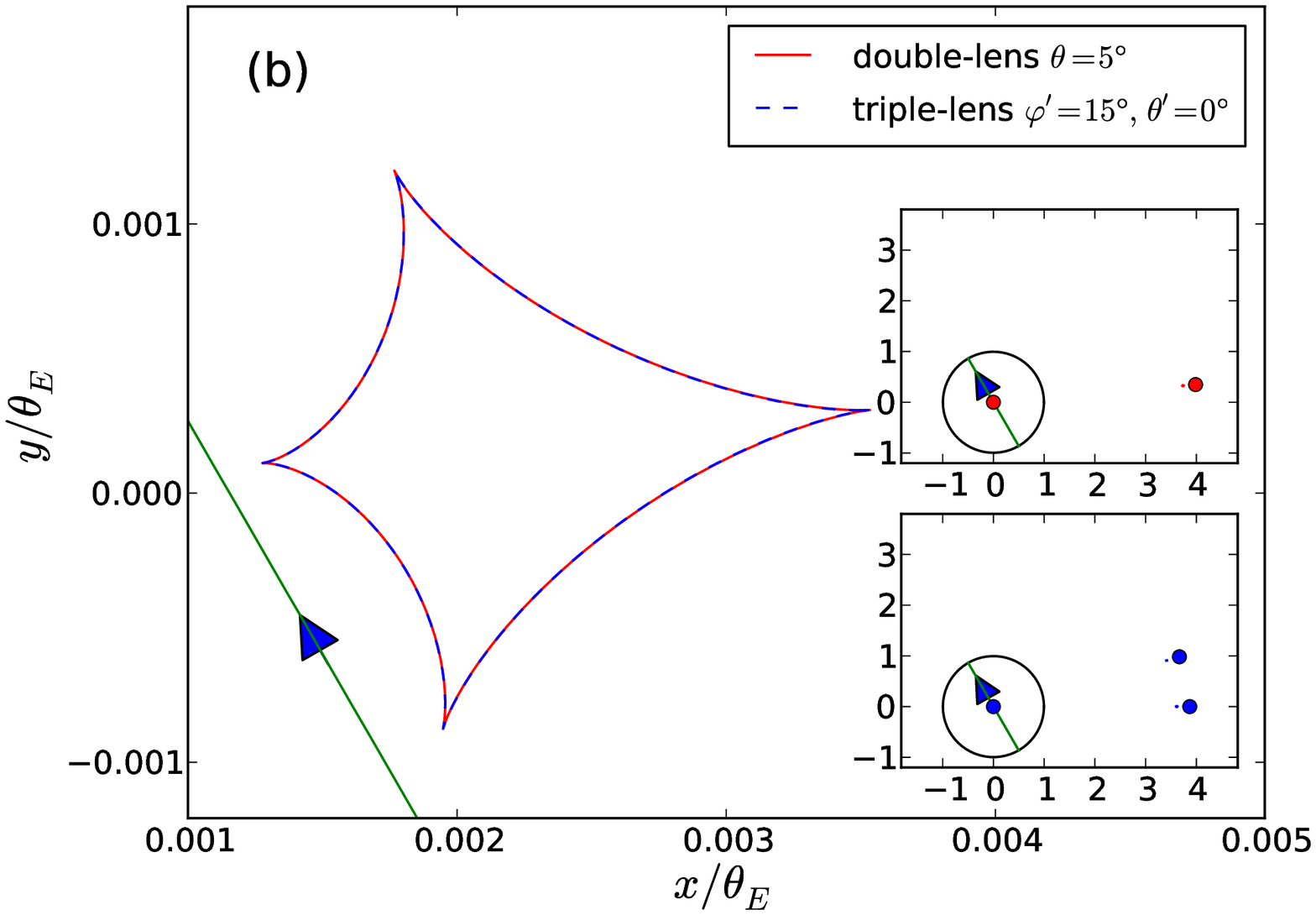}
\includegraphics[width=.45\textwidth]{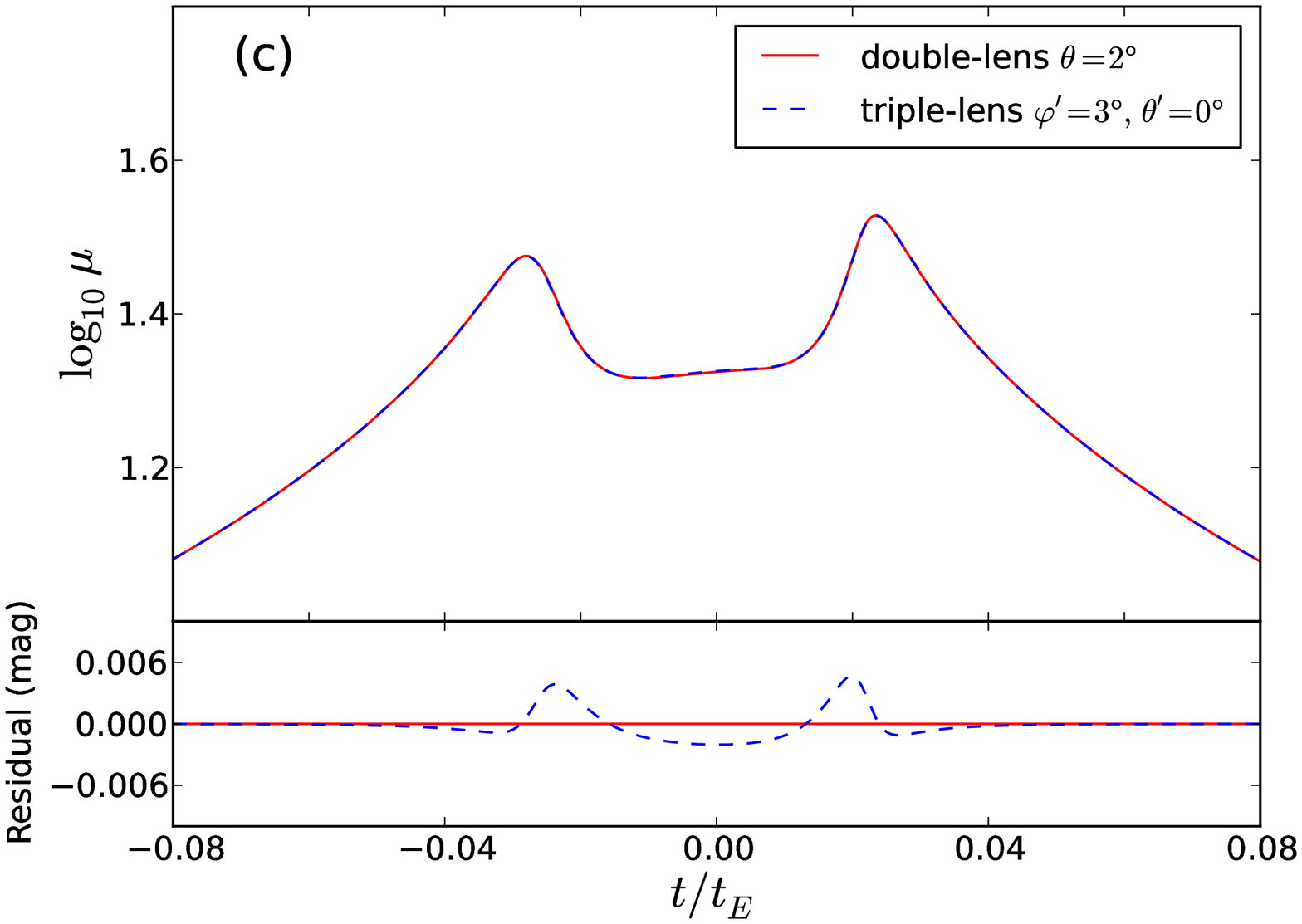}
\includegraphics[width=.45\textwidth]{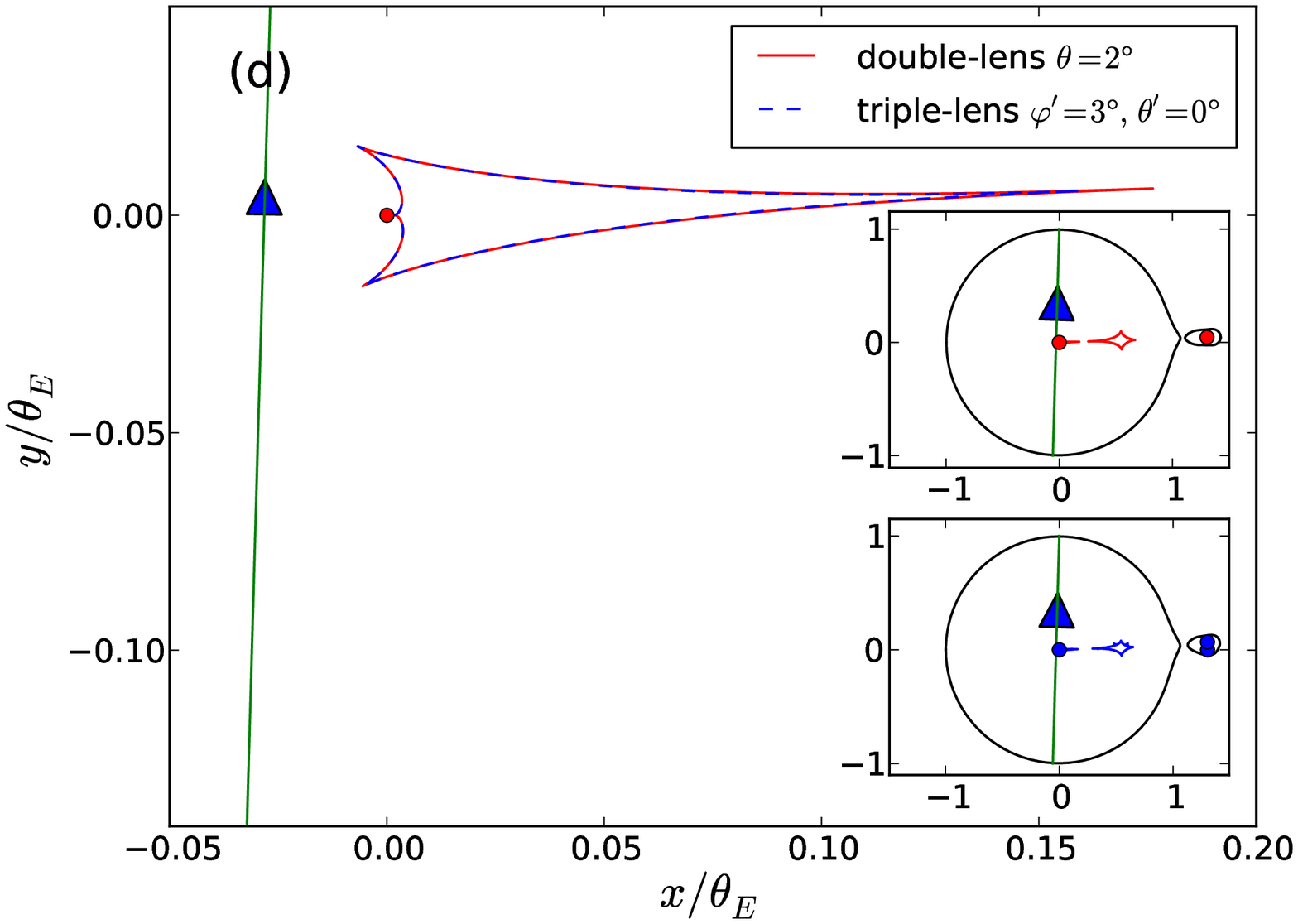}
\includegraphics[width=.45\textwidth]{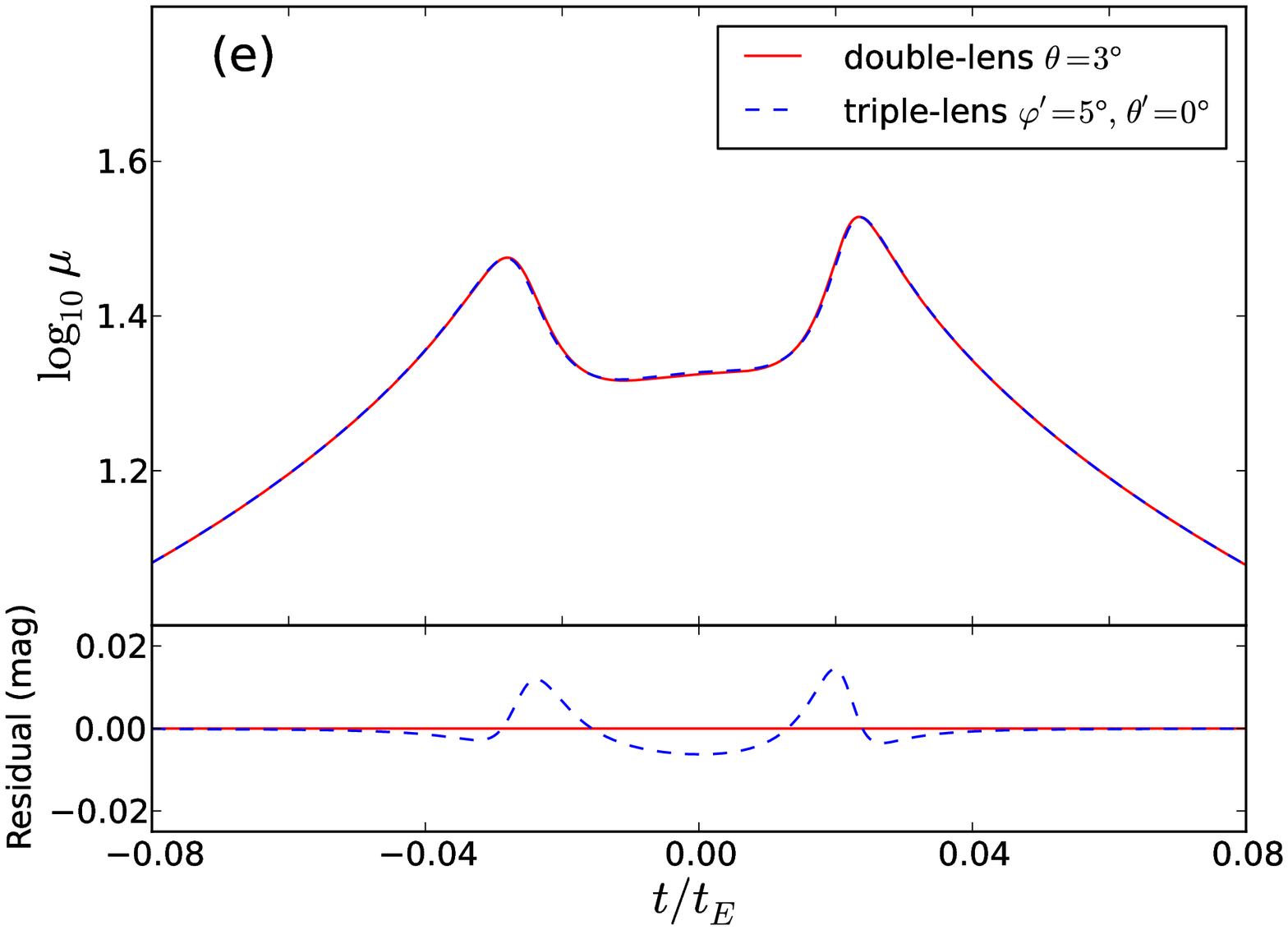}
\includegraphics[width=.45\textwidth]{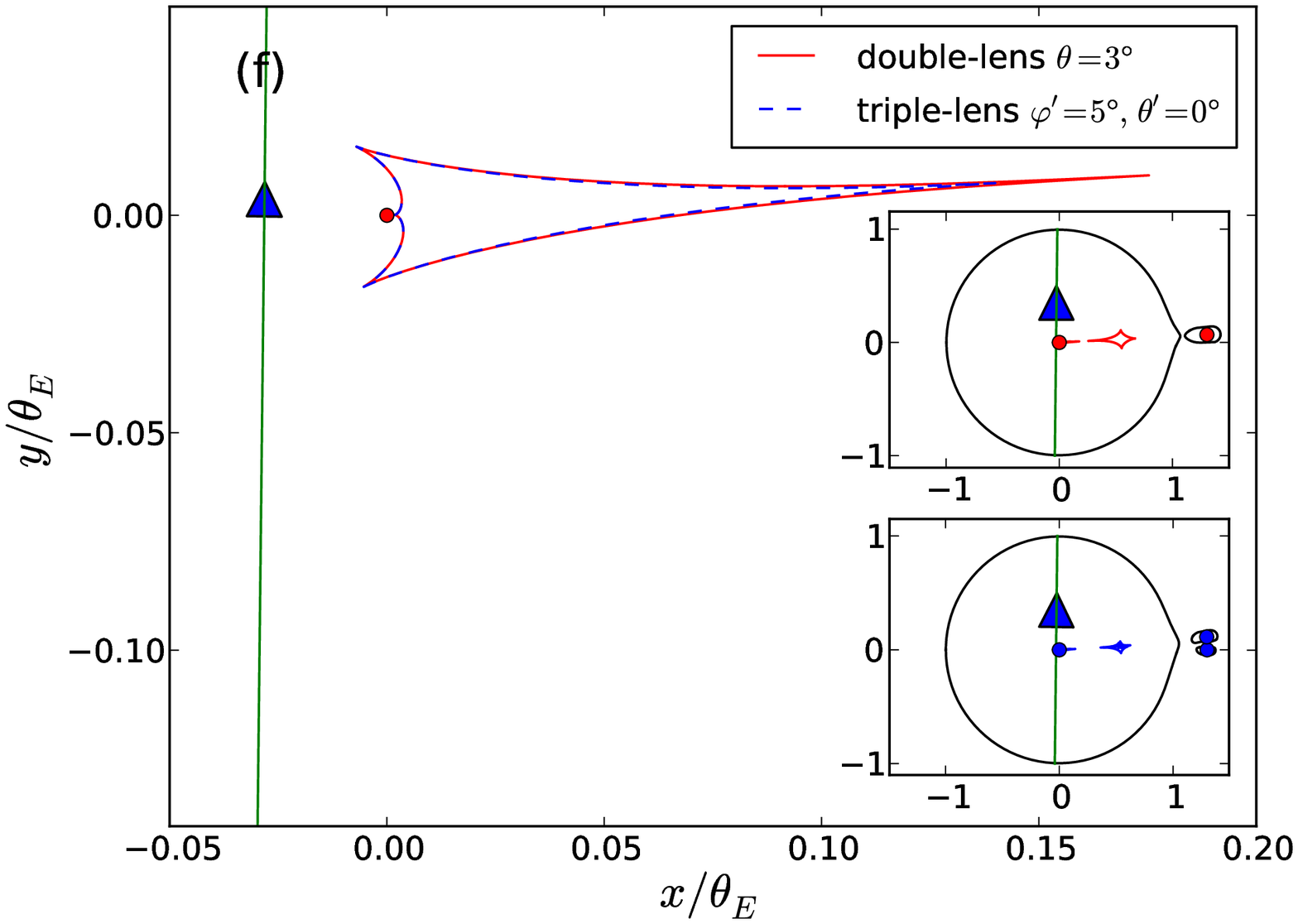}
\caption{Examples of the double-triple lens degeneracy from double-lens to triple-lens. The left panel shows peak-region light curves with the residual between them at the bottom; the right panel shows the central caustics with the overall configurations in the insets. The red solid lines represent the initial static double-lens system, and the blue dashed lines are for the derived triple-lens system. For the insets, the straight green lines with an arrow are the trajectories, the black rounded curves are the critical curves, the colored dots are the lenses and the colored curves are the caustics (which may be too small to see). The examples in the middle and bottom panels are taken from OGLE-2005-BLG-071 \citep{Dong09} with a recalculation shown in Appendix~\ref{app3}. All the parameters are shown in Table~\ref{tab:para}(3).}
\label{fig:double2}
\end{figure*}

\begin{figure*}
\graphicspath{}
\centering
\vspace{-3mm}
\includegraphics[width=.33\textwidth]{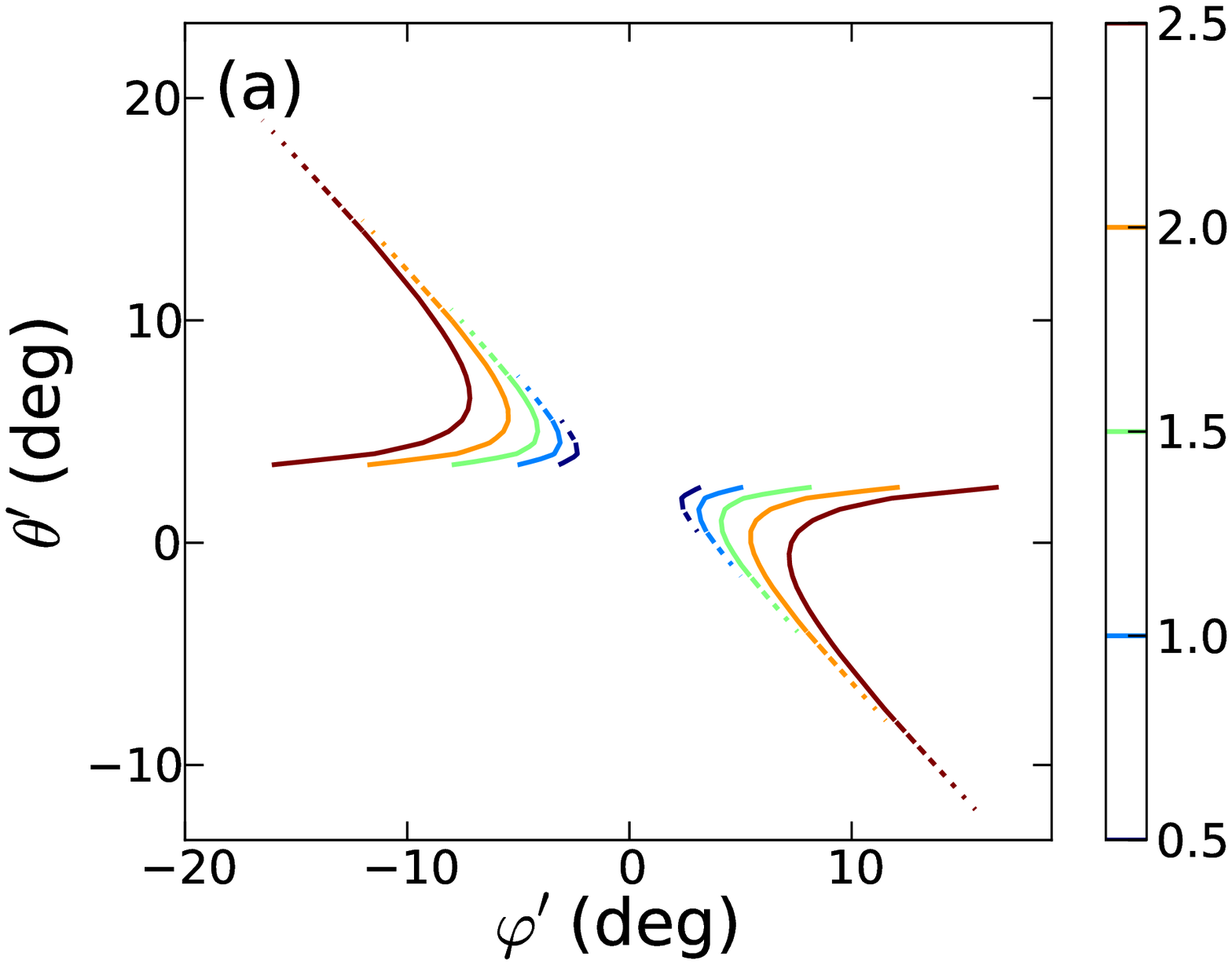}
\includegraphics[width=.33\textwidth]{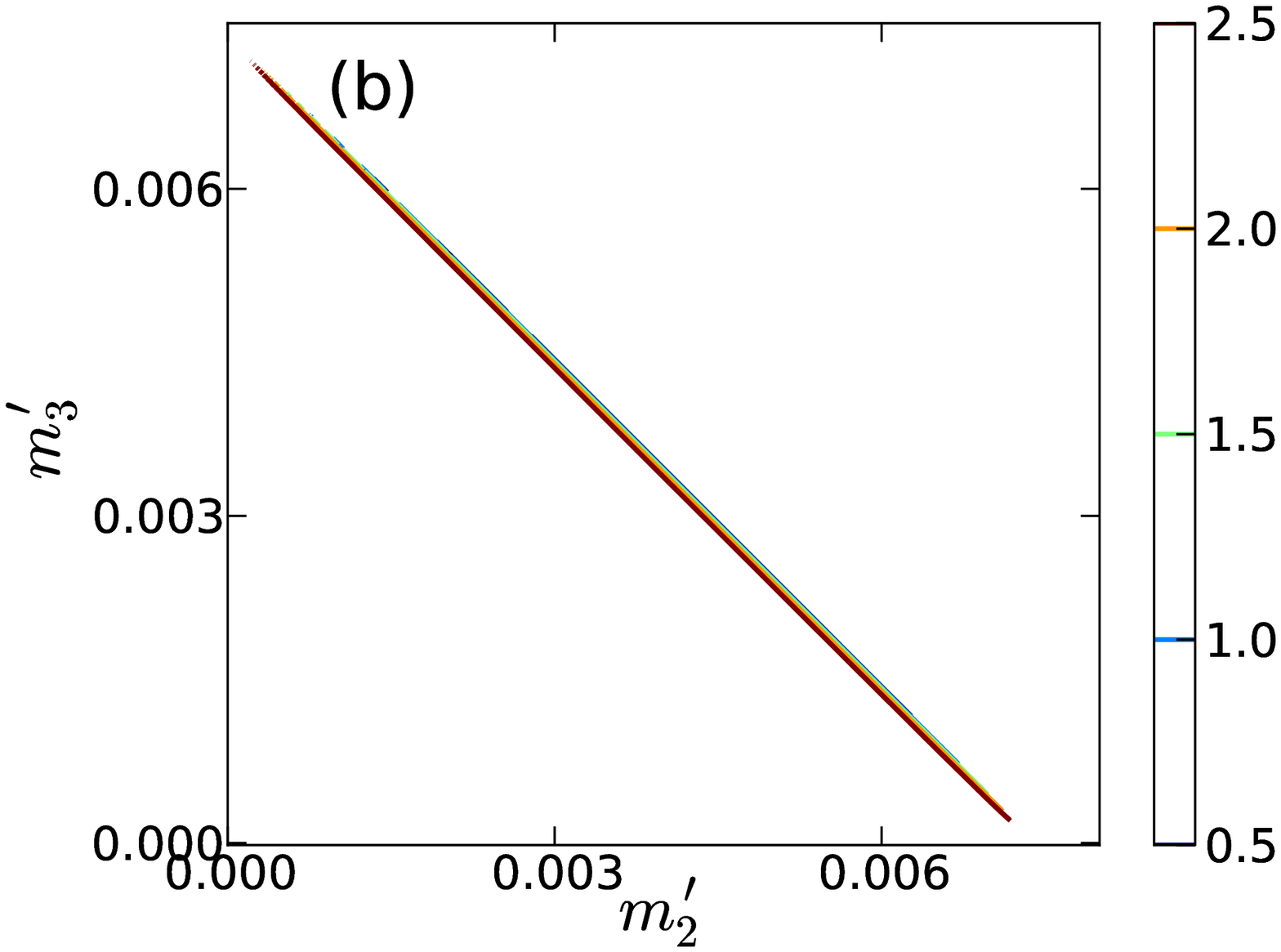}
\includegraphics[width=.33\textwidth]{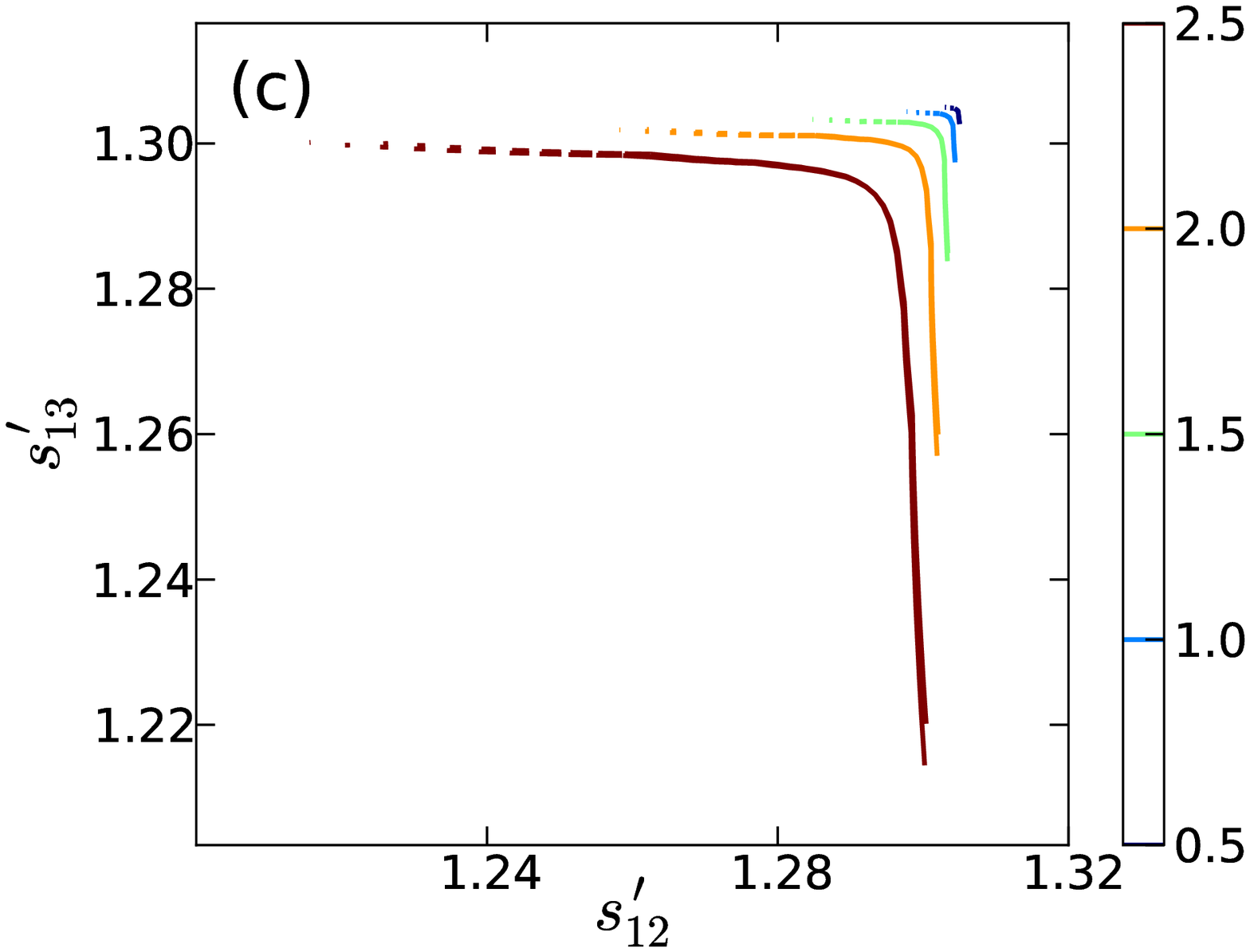}
\includegraphics[width=.33\textwidth]{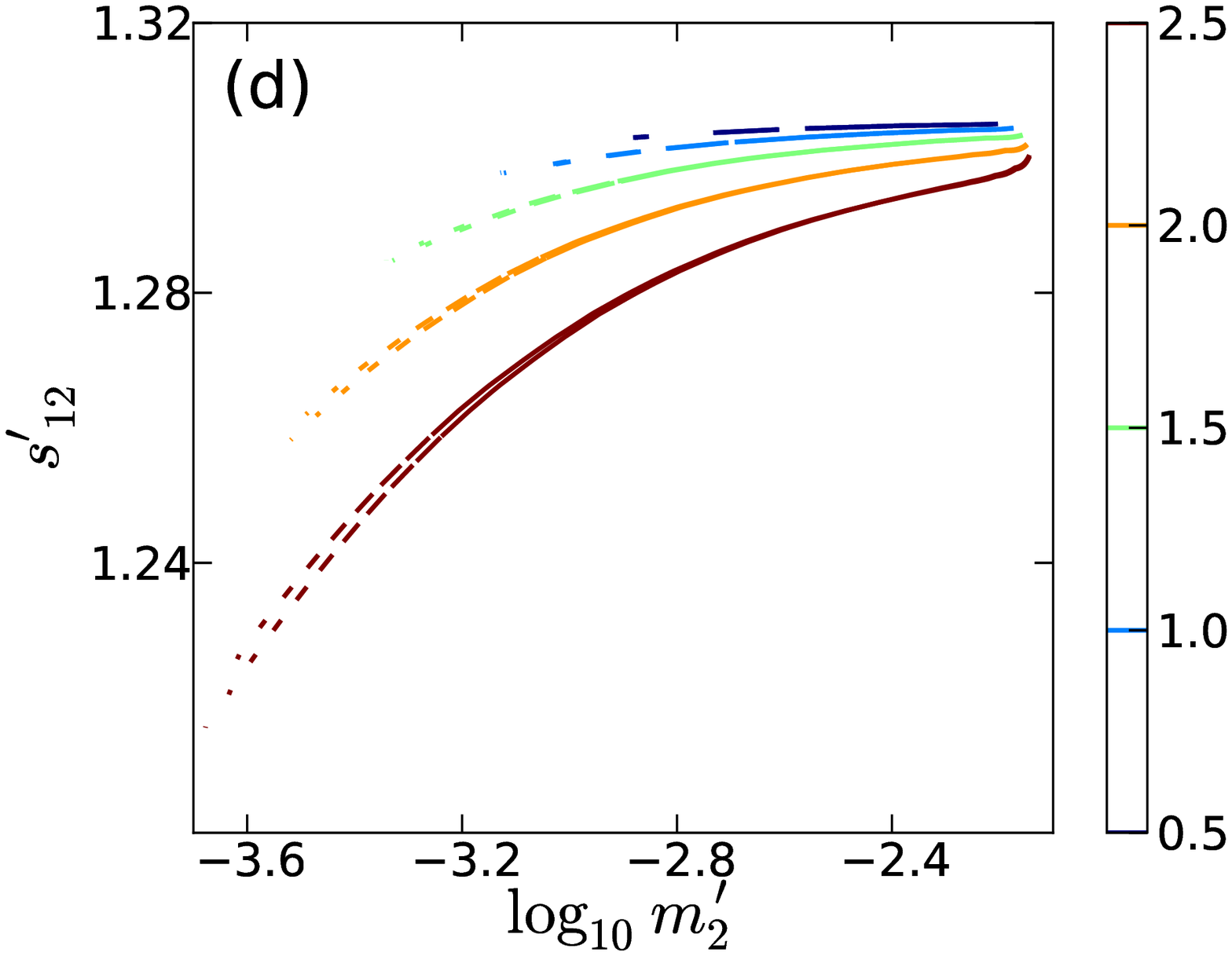}
\includegraphics[width=.33\textwidth]{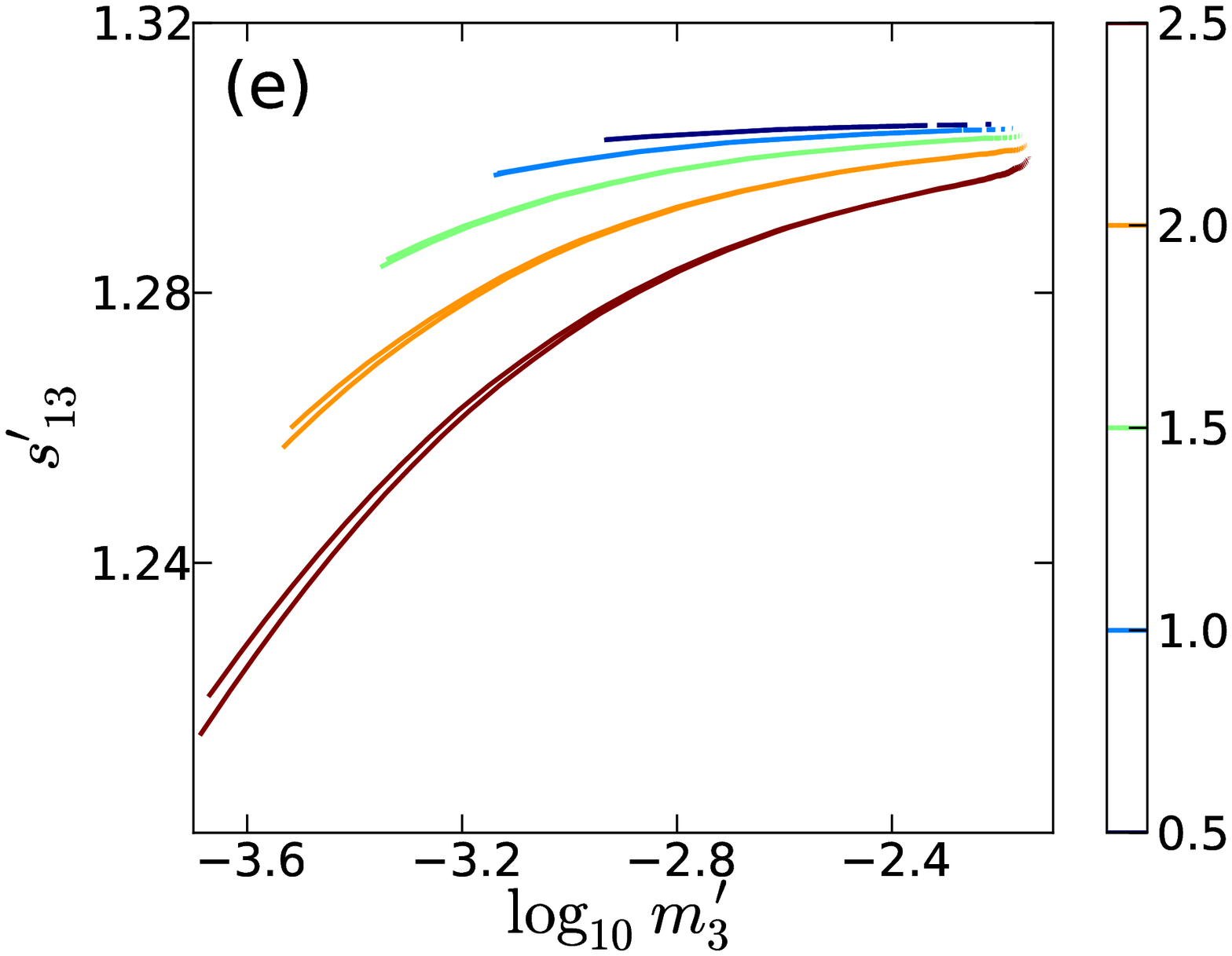}
\caption{Contours of $\Delta\chi^2$ between two parameters of the triple-lens system used in Fig.~\ref{fig:double2}(c). The symbols are the same as in Fig. \ref{fig:c30}. The parameters follow roughly a single parameter family as discussed in \S\ref{sec:double} and appendix \ref{subsec:2-3}.}
\label{fig:c23}
\end{figure*}

\begin{figure*}
\graphicspath{}
\centering
\vspace{-3mm}
\includegraphics[width=.45\textwidth]{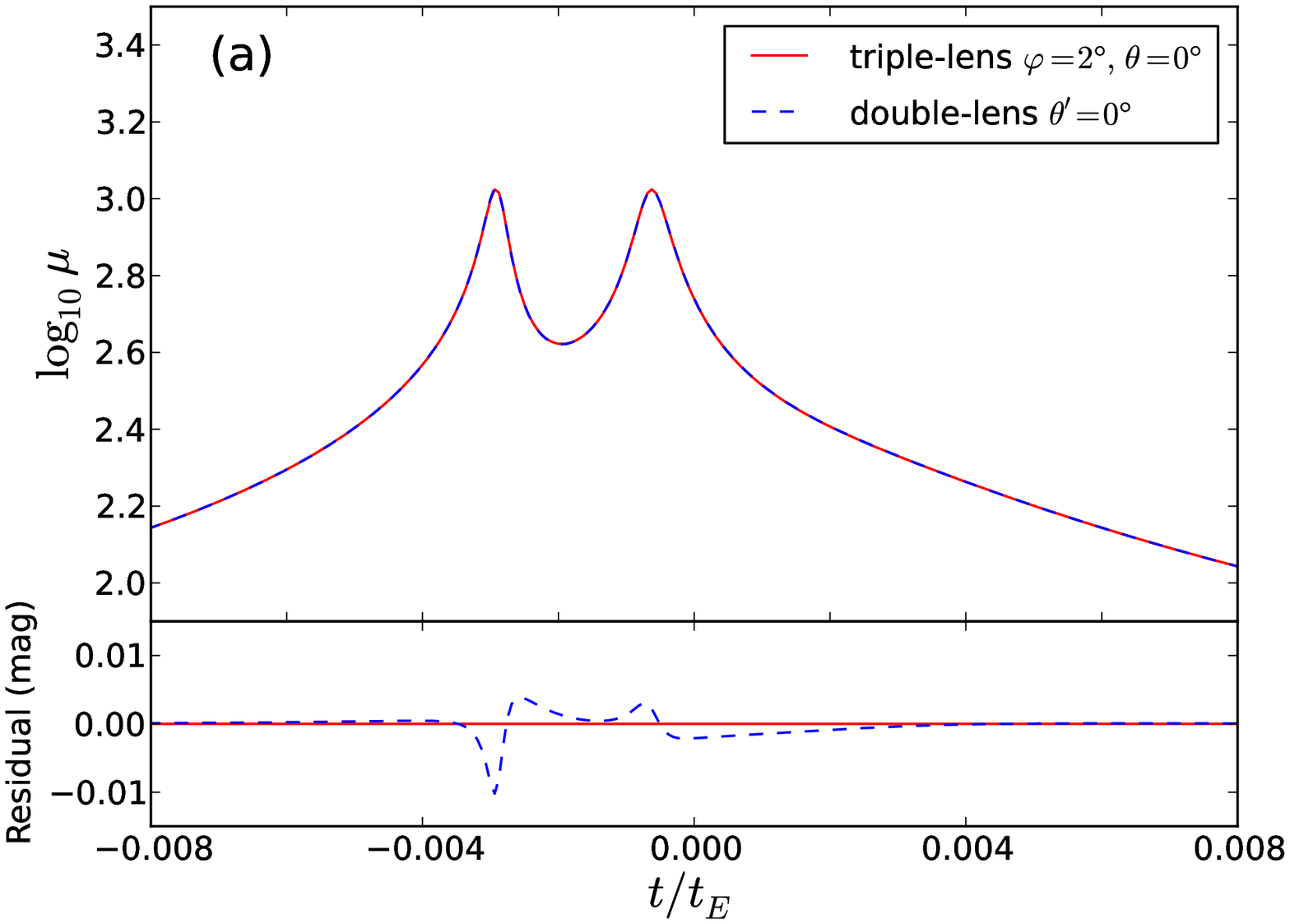}
\includegraphics[width=.45\textwidth]{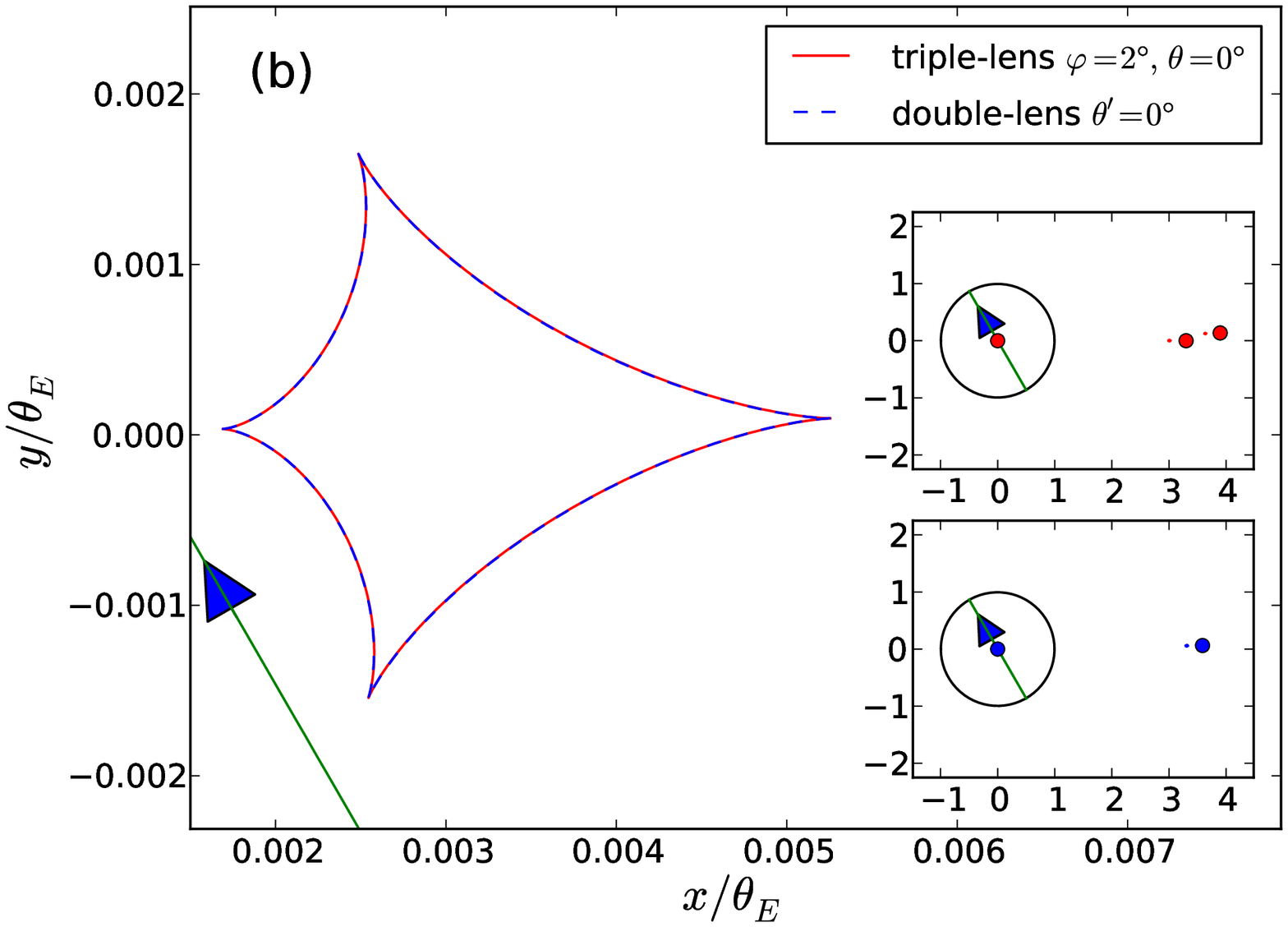}
\includegraphics[width=.45\textwidth]{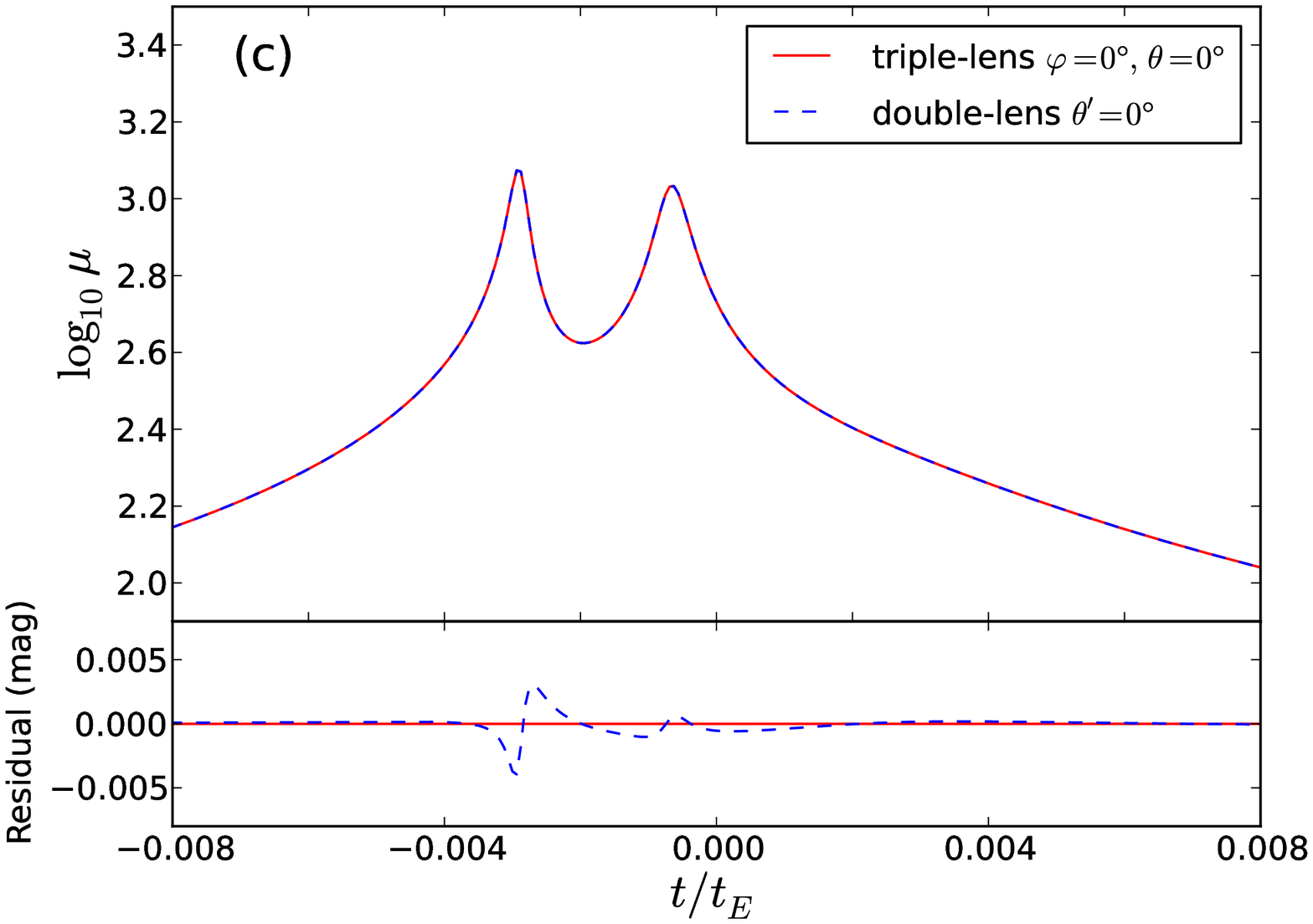}
\includegraphics[width=.45\textwidth]{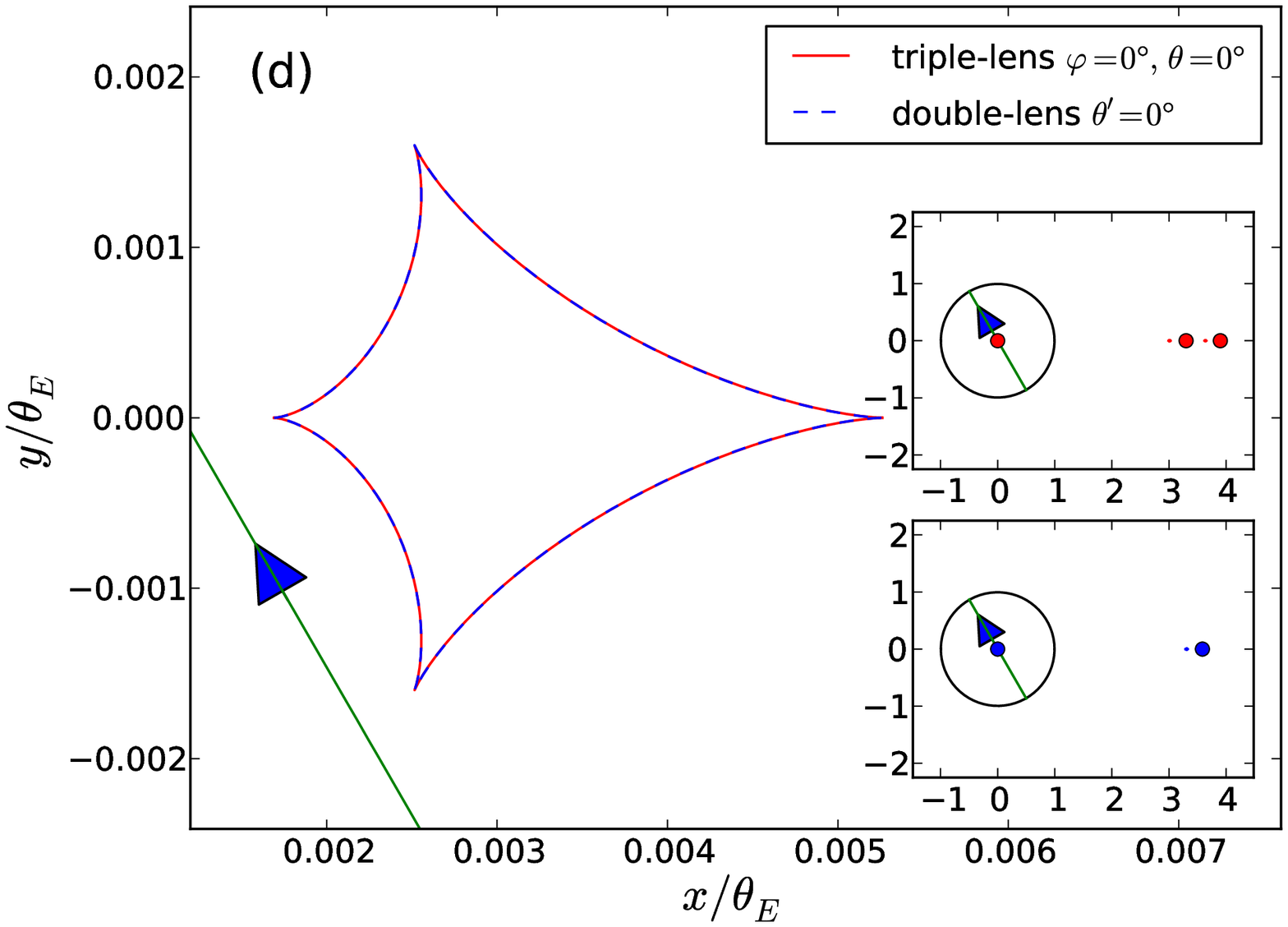}
\caption{Examples of the double-triple lens degeneracy from triple-lens to double-lens. The left panel shows peak-region light curves with the residual between them at the bottom; the right panel shows the central caustics with the overall configurations in the insets. The red solid lines represent the initial static triple-lens system, and the blue dashed lines are for the derived double-lens system. For the insets, the straight green lines with an arrow are the trajectories, the black rounded curves are the critical curves, the colored dots are the lenses and the colored curves are the caustics (which may be too small to see). All the parameters are shown in Table~\ref{tab:para}(4).}
\label{fig:double1}
\end{figure*}

As mentioned in the last subsection, we find a ``double-triple lens degeneracy'' which we now discuss in greater detail (see also \citealt{Gau98}).

By setting $m_3 = 0$, equations~(\ref{Taylor})~to~(\ref{shear}) in \S\ref{sec:shear} recover a static double-lens system with the parameters $(m_2,\,s,\,\theta)$, and hence we can write the four external shear equations for double-lens as
\begin{subequations}
\label{2shearEq}
\begin{align} 
 {m_2 \over s^2} \cos{2\theta} & \equiv a_1, \\
 {m_2 \over s^2} \sin{2\theta} & \equiv b_1, \\
 {m_2 \over s^3} \cos{3\theta} & \equiv {a_2 \over \sqrt{m_1}}, \\
 {m_2 \over s^3} \sin{3\theta} & \equiv {b_2 \over \sqrt{m_1}}. 
\end{align}
\end{subequations}
Then the degenerate parameters can be calculated by the same analytic method mentioned in \S\ref{sec:shear} (see also Appendix~\ref{subsec:3-3}).

In principle, there are two ways to apply our degeneracy solution depending on which kind of systems is the initial one. We will discuss these in turn.
\begin{enumerate}
\item The first is that we have a double-lens system known and need to find a degenerate triple-lens system with subtle residuals. In practice, this application is important since it is natural for modellers to fit binary-lens models first rather than the more complex triple lenses.

To do this, we first use $(m_2,\,s,\,\theta)$ to calculate $(a_1,\,b_1,\,a_2,\,b_2)$ through equations~(\ref{2shearEq}). Note that, if ${\alpha}_0 \rightarrow 0^\circ$ or $180^\circ$, $b_1\approx0$ and $b_2\approx0$, the method would fail. Afterwards, with certain $(\varphi',\,\theta')$, the remaining four derived triple-lens parameters $(m_2',\,m_3',\,s_{12}',\,s_{13}')$ can be determined by equations~(\ref{3shearEq}) (see Appendix~{\ref{subsec:2-3}} for the detailed procedure). Usually, the best degenerate positions of $m_2$ and $m_3$ in a derived triple-lens system are always at the vicinity of the $m_2$ positions in the initial double-lens system. Through this procedure, many ``continuous" degenerate solutions may be found.

Fig.~\ref{fig:double2} gives three examples of the double-triple lens degeneracy from double-lens to triple-lens. The red solid lines represent the initial double-lens system, and the blue dashed lines are for one of the possible degenerate triple-lens system. As mentioned before, a shift in the source position is needed to draw the central caustics together (equation~\ref{shift}). The top panel shows an artificial example with $\varphi'=15^\circ$ and $\theta'=0^\circ$. For the middle and bottom panels, the initial binary lens parameters are taken from a real confirmed planetary microlensing event OGLE-2005-BLG-071 (\citealt{Ud05, Dong09}). The input parameters are given in Table \ref{tab:para}(3). By the comparison of these two last panels, one can find that the degeneracy is stronger with smaller $\varphi'$.

Fig.~\ref{fig:c23} shows the correlations between the derived parameters which are calculated by the parameters of OGLE-2005-BLG-071 using the same cadence and error bar models as in \S\ref{sec:triple}. As can be seen, in this case, the $\Delta\chi^2$ surface follows roughly straight lines. We show in the Appendix \ref{subsec:2-3} that this can be understood quite easily since the solutions can be roughly expressed as a one-parameter family.
\item
The second is opposite to the procedure mentioned above, i.e., we have found a triple-lens system and then want to explore the degeneracy due to possible double-lens systems. In this case, we can use a triple-lens system with $(m_2,\,m_3,\,s_{12},\,s_{13},\,\varphi,\,\theta)$ to determine $(a_1,\,b_1,\,a_2,\,b_2)$ through equations~(\ref{3shearEq}). And then $(m_2',\,s',\,\theta')$ can be calculated via equations~(\ref{2shearEq}) (see Appendix~{\ref{subsec:3-2}} for the detailed procedure). 

Specially, if the derived direction angle $\theta' = 0^\circ$, equations~(\ref{2shearEq}b) and (\ref{2shearEq}d) will be equal to 0, which demand that the initial triple-lens system should satisfy $b_1\approx0$ and $b_2\approx0$. As a result, the feasible initial angular parameters $\varphi$ and $\theta$ are limited to near either $0^\circ$ or $180^\circ$ respectively, and $\Delta\theta \approx 0^\circ$. 

Fig.~\ref{fig:double1} gives two examples of the double-triple lens degeneracy from triple-lens to double-lens. The red solid lines represent the initial triple-lens system, while the blue dashed lines are for one of the possible degenerate double-lens system. The parameters are given in Table \ref{tab:para}(4). The differences in both examples are very small ($\sim 0.005$ mag). 
\end{enumerate}

\section{Other extreme triple-lens systems}
\label{sec:other}
In \S\ref{sec:shear}, we discussed the extreme case when all the other lenses are much farther away from the primary lens. There are two more extreme cases: one is when two lenses are close to each other with the last lens located far away, and the other is when all three lenses are close together. Note that the three cases have been pointed out by \cite{Dom99} for binary-lens systems and were re-examined by \cite{Boz00} for multiple-lens systems. We mention them here by using a different mathematical method for completeness, but shall explore them elsewhere.

\subsection{Close pair plus one wide companion}
According to \S\ref{sec:shear}, we can expand the deflection term in series when its corresponding lens is far from the primary lens. Similarly, if two lenses are close, we can handle the deflection term by means of the multipole expansion (\citealt{Dom99, An05}). So for a close pair with one wide companion, the lens equation may be series-expanded as
\label{sec:close-pair}
\begin{align}  
\label{close-pair}
 \zeta = & z - {m_1 \over \bz - \bz_1} - {m_2 \over \bz - \bz_2} - {m_3 \over \bz - \bz_3}\nonumber\\
       = & z-{m_1 + m_2 \over \bz - \bzc} - {m_1\left(\bz_1-\bzc\right) + m_2\left(\bz_2-\bzc\right) \over \left(\bz-\bzc\right)^2} \nonumber\\
         & - \sum_{k=2}^{\infty} { m_1\left(\bz_1-\bzc\right)^{k} + m_2\left(\bz_2-\bzc\right)^{k} \over \left(\bz-\bzc\right)^{k+1}} \nonumber \\
         & + {m_3 \over \bz_3 - \bzc} + \sum_{k=1}^{\infty} {{m_3 \over \left(\bz_3 - \bzc\right)^{k+1}} \left(\bz - \bzc\right)^{k}},
\end{align}
provided $|\bz - \bzc| \gg \max \left(|\bz_1 - \bzc|,\,|\bz_2 - \bzc|\right)$ and $|\bz - \bzc| \ll |\bz_3 - \bzc|$. If $z_{\rm c}$ is chosen to be the centre of mass of the close pair
\begin{equation}
 z_{\rm c} = {m_1 z_1 + m_2 z_2\over m_1 + m_2},
\end{equation}
then the dipole term [$\alpha(\bz-\bzc)^{-2}$] in equation~(\ref{close-pair}) vanishes. Next, we can transform equation~(\ref{close-pair}) into
\begin{equation}
 \omega = w - {1 \over \bar{w}} - \sum_{k=2}^{\infty} { {Q^k} \over \bar{w}^{k+1}}
           + \sum_{k=0}^{\infty} { {\gamma_k} \bar{w}^{k+1}},
\end{equation}
where
\begin{gather}
 \omega = {{\zeta - \zeta_{\rm c}} \over \sqrt{m_1+m_2}}, \qquad w = {{z - z_{\rm c}} \over \sqrt{m_1+m_2}}, \\
 \zeta_{\rm c} = z_{\rm c} + {m_3 \over {\bz_3 - \bzc}}, \\
 \begin{align}
   Q^k = & {m_1\left(\bz_1-\bzc\right)^{k} + m_2\left(\bz_2-\bzc\right)^{k} \over  \left(m_1+m_2\right)^{{k \over 2} + 1}} \nonumber \\
       = & {m_1 m_2 \left[m_2^{k-1} - (-m_1)^{k-1}\right] \left(\bz_1 - \bz_2\right)^{k} \over \left(m_1+m_2\right)^{{3k \over 2} + 1}},
 \end{align}
 \\ 
 \gamma_k = {\left(m_1+m_2\right)^{k \over 2} m_3 \over \left(\bz_3 - \bzc\right)^{k+2}} 
          = {\left(m_1+m_2\right)^{{3k \over 2}+1} m_3 \over \left[m_1\left(\bz_3 - \bz_1\right)+ m_2\left(\bz_3 - \bz_2\right)\right]^{k+2}}.
\end{gather}

\subsection{Close triple lens}
\label{sec:close-triple}
Similarly, when all three lens are close, the multipole expansion of the lens equation results in
\begin{align}  
\label{close-triple}
 \zeta = & z - {m_1 \over \bz - \bz_1} - {m_2 \over \bz - \bz_2} - {m_3 \over \bz - \bz_3}\nonumber\\
       = & z-{m \over \bz - \bzc} - \sum_{i=1}^3 {m_i\left(\bz_i-\bzc\right) \over \left(\bz-\bzc\right)^2} - \sum_{k=2}^{\infty} \sum_{i=1}^3 { m_i\left(\bz_i-\bzc\right)^{k} \over \left(\bz-\bzc\right)^{k+1}},
\end{align}
where $m = m_1 + m_2 + m_3\ (\equiv 1)$, provided $|\bz - \bzc| \gg \max \left(|\bz_1 - \bzc|,\,|\bz_2 - \bzc|,\,|\bz_3 - \bzc|\right)$. As before, if we choose $z_{\rm c}$ to be the centre of mass of the close triple system
\begin{equation}
 z_{\rm c} = \sum_{i=1}^3 {m_i z_i \over m},
\end{equation}
then the dipole term in equations~(\ref{close-triple}) vanishes. Finally, we have
\begin{equation}
 \omega = w - {1 \over \bar{w}} - \sum_{k=2}^{\infty} { {Q^k} \over \bar{w}^{k+1}},
\end{equation}
where
\begin{gather}
 \omega = {{\zeta - z_{\rm c}} \over \sqrt{m}}, \qquad w = {{z - z_{\rm c}} \over \sqrt{m}}, \\
 Q^k = \sum_{i=1}^3 {m_i\left(\bz_i-\bzc\right)^{k} \over m^{{k \over 2} + 1}}.
\end{gather}

\section{Discussion}
\label{sec:discussion}

In this paper, we have studied the degeneracies in triple gravitational microlensing. First of all, a discrete degeneracy is obvious by reversing the sign of the three parameters $(u_0,\,\alpha_0,\,\varphi)$, and it can be broken when parallax effects are considered. Secondly, a four-fold close/wide degeneracy is derived mathematically for a planetary system with two planets which is consistent with the conclusion drawn by \cite{Boz00}. Thirdly, a continuous external shear degeneracy is confirmed to exist either among many different triple-lens systems or between double-lens systems and triple-lens systems (mentioned in passing by \citealt{Gau98}.) Finally we mentioned but not explored two other extreme case of triple lensing (\S\ref{sec:other}).

We also give detailed recipes to calculate the parameters satisfying the external shear degeneracy (see Appendix~\ref{app2}). With these recipes, the whole parameter space can be searched through numerical method, e.g., Monte Carlo Markov Chain method.

The continuous degeneracy implies that the double and triple lenses may be degenerate. This has the important consequence that in some cases, multiple planet systems may be mistakenly identified as a single planet system. If this happens, a wrong set of planetary parameters may be derived and the frequency of multiple planet systems will be under-estimated.

Naively the probability for triple lensing may be somewhat lower due to binary lensing since it requires two planets to be present. However, if all the systems are in a single orbital plane, then the chance of detecting two planets may be boosted when viewed edge on. The probability of this degeneracy being observed will depend on the detailed predictions from the planet formation theories. 

Indeed microlensing is perhaps the only way to probe multiple planet population at a few AU. The next-generation microlensing experiment such as the Korean Microlensing Telescope Network (KMTNet) presents an exciting possibility to explore this parameter space.

\section*{Acknowledgments}

We thank Subo Dong and Andy Gould for helpful discussions. We acknowledge the Chinese Academy of Sciences and National Astronomical Observatories of China for financial support and the Institute of Astronomy at Cambridge for hospitality and the chilly British summer in 2012.

\bibliographystyle{mn2e}

\appendix

\section{Analytic method for the external shear equations}
\label{app2}

\subsection{The external shear degeneracy}
\label{subsec:3-3}

In the first place, it is crucial to eliminate $\sqrt{m_1}$ in equations~(\ref{3shearEq}), because the initial and derived lens systems might have different $m_1$ which will make the calculation more complicated. By defining
\begin{equation}
\label{substiEq}
 q_{21} = {m_2 \over m_1}, \quad q_{31} = {m_3 \over m_1}, \quad d_{12} = {s_{12} \over \sqrt{m_1}}, \quad d_{13} = {s_{13} \over \sqrt{m_1}},
\end{equation}
we can rewrite equations~(\ref{3shearEq}) as
\begin{subequations}
\label{initialEq}
\begin{align}
 {q_{21} \over d_{12}^2} \cos{2\theta} + {q_{31} \over d_{13}^2} \cos{2\left(\varphi+\theta\right)} & \equiv a_1, \\
 {q_{21} \over d_{12}^2} \sin{2\theta} + {q_{31} \over d_{13}^2} \sin{2\left(\varphi+\theta\right)} & \equiv b_1, \\
 {q_{21} \over d_{12}^3} \cos{3\theta} + {q_{31} \over d_{13}^3} \cos{3\left(\varphi+\theta\right)} & \equiv a_2, \\
 {q_{21} \over d_{12}^3} \sin{3\theta} + {q_{31} \over d_{13}^3} \sin{3\left(\varphi+\theta\right)} & \equiv b_2.
\end{align}
\end{subequations}

With the initial triple-lens parameters $(q_{21},\,q_{31},\,d_{12},\,d_{13},\,\varphi,\,\theta)$, one can calculate the constants $(a_1,\,b_1,\,a_2,\,b_2)$ through equations~(\ref{initialEq}).

The next step is to find other sets of lens parameters $(q_{21}',\,q_{31}',\,d_{12}',\,d_{13}',\,\varphi',\,\theta')$ satisfying equations~(\ref{initialEq}). This can be achieved analytically by choosing $(\varphi',\,\theta')$ as the remaining free parameters, and thus $(q_{21}',\,q_{31}',\,d_{12}',\,d_{13}')$ are represented as a function of $(\varphi',\,\theta')$
\begin{subequations}
\label{solutions}
\begin{align}
 d_{12}' &= {A_1 \sin{3\varphi'} \over A_2 \sin{2\varphi'}}, \\
 q_{21}' &= {A_1 d_{12}'^2 \over \sin{2\varphi'}}, &&\\
 d_{13}' &= {B_1 \sin{3\varphi'} \over B_2 \sin{2\varphi'}}, \\
 q_{31}' &= {B_1 d_{13}'^2 \over \sin{2\varphi'}}, \\
 m_1' &= {1 \over {1+q_{12}'+q_{13}'}},
\end{align}
\end{subequations}
where 
\begin{subequations}
\label{coeff}
\begin{align}
 &A_1 = a_1 \sin{2\left(\varphi'+\theta'\right)} - b_1 \cos{2\left(\varphi'+\theta'\right)}, \\
 &A_2 = a_2 \sin{3\left(\varphi'+\theta'\right)} - b_2 \cos{3\left(\varphi'+\theta'\right)}, \\
 &B_1 = b_1 \cos{2\theta'} - a_1 \sin{2\theta'}, \\
 &B_2 = b_2 \cos{3\theta'} - a_2 \sin{3\theta'}. 
\end{align}
\end{subequations}
Note that, equations~(\ref{substiEq}) are still needed to obtain $(m_2',\,m_3',\,s_{12}',\,s_{13}')$.

Obviously, there are some special $(\varphi',\,\theta')$ which will make the analytic method invalid: (1) $\sin{2\varphi'} = 0$, i.e., $\varphi' = 0^\circ$, $90^\circ$ or $180^\circ$; (2) $\sin{3\varphi'} = 0$, i.e., $\varphi' = 60^\circ$ or $120^\circ$; and (3) $A_1$ or $A_2$ or $B_1$ or $B_2 = 0$. See \S\ref{sec:triple} for more discussions about the initial parameters.

\subsection{The double-triple lens degeneracy: from triple-lens to double-lens}
\label{subsec:3-2}
In this case, starting with an initial triple-lens system $(m_2,\,m_3,\,s_{12},\,s_{13},\,\varphi,\,\theta)$, we need to calculate a degenerate double-lens system $(m_2',\,s',\,\theta')$. A substitution is still required before the calculation
\begin{equation}
\label{substiEq2}
 q = {m_2 \over m_1}, \quad d = {s \over \sqrt{m_1}},
\end{equation}
so equations~(\ref{2shearEq}) become
\begin{subequations}
\label{2shear1}
\begin{align}
 {q \over d^2} \cos{2\theta} & = a_1, \\
 {q \over d^2} \sin{2\theta} & = b_1, \\
 {q \over d^3} \cos{3\theta} & = a_2, \\
 {q \over d^3} \sin{3\theta} & = b_2, 
\end{align}
\end{subequations}
where the constants $(a_1,\,b_1,\,a_2,\,b_2)$ should be calculated by equations~(\ref{initialEq}).

The analytic solutions are
\begin{subequations}
\label{solutions2}
\begin{align}
 &d' = \sqrt{{a_1^2+b_1^2} \over {a_2^2+b_2^2}}, \\
 &q' = {d'^2} \sqrt{a_1^2+b_1^2}, \\
 &m_1' = {1 \over {1+q'}},
\end{align}
\end{subequations}
with $\theta'$ satisfying
\begin{subequations}
\label{alpha}
\begin{align}
 \tan{2\theta'} & \approx {b_1 \over a_1}, \\
 \tan{3\theta'} & \approx {b_2 \over a_2},
\end{align}
\end{subequations}
at the same time. One simplification is to set either $\theta'=0^\circ$ or $\theta=0^\circ$ in the calculation, although some of the solutions might be lost. 

One special case is when the masses of the initial triple-lens system are located on a line, i.e., $\varphi=0^\circ$ or $180^\circ$, this method will always work with $\theta'-\theta=0^\circ$ or $180^\circ$ (e.g., Fig.~\ref{fig:double1}c and \ref{fig:double1}d).

\subsection{The double-triple lens degeneracy: from double-lens to triple-lens}
\label{subsec:2-3}
In this case, we initially have a double-lens system $(m_2,\,s,\,\theta)$ and need to calculate a degenerate triple-lens system $(m_2',\,m_3',\,s_{12}',\,s_{13}',\,\varphi',\,\theta')$. Now, the constants $(a_1,\,b_1,\,a_2,\,b_2)$ should be calculated by equations~(\ref{2shear1}), and then equations~(\ref{initialEq}) are called again to obtain the derived parameters. As a result, the analytic solutions are the same as equations~(\ref{solutions}). 

Note that, equations~(\ref{coeff}) can be simplified by equations~(\ref{2shear1}) into
\begin{subequations}
\begin{align}
 &A_1 = {q \over d^2} \sin{2\left(\varphi'+\theta'-\theta\right)}, \\
 &A_2 = {q \over d^3} \sin{3\left(\varphi'+\theta'-\theta\right)}, \\
 &B_1 = {q \over d^2} \sin{2\left(\theta'-\theta\right)}, \\
 &B_2 = {q \over d^3} \sin{3\left(\theta'-\theta\right)}. 
\end{align}
\end{subequations}
So, it is easy to derive the relation between $q_{12}'$ and $q_{13}'$, i.e.,
\begin{equation}
\label{mass-relation}
 q_{12}'+q_{13}' = q {\sin^2{3\varphi'} \over \sin^3{2\varphi'}} \left[{\sin^3{2\left(\varphi'+\theta'-\theta\right)} \over \sin^2{3\left(\varphi'+\theta'-\theta\right)}} + {\sin^3{2\left(\theta'-\theta\right)} \over \sin^2{3\left(\theta'-\theta\right)}} \right],
\end{equation}
which explains the correlation in Fig.~\ref{fig:c23}(b): if $\theta'=\theta$, then the degeneracy is described by a single parameter $\varphi'$, only.

\section{The lens parameters of the examples}
\label{app3}
The lens parameters used in Figs.~\ref{fig:4-fold}, \ref{fig:triple}, \ref{fig:double2} and \ref{fig:double1} are shown in Table~\ref{tab:para}. 

For the last two examples in Fig.~\ref{fig:double2}, we use the set of parameters in the ``Wide+'' case of ``MCMC A'' from \cite{Dong09}
\begin{equation}
 u_0 = 0.0282, \quad d = 1.306, \quad q = 7.5 \times 10^{-3}, \quad \alpha = 273.63^\circ.
\end{equation}
In our notations, $(u_0,\,d)$ would remain the same regardless the slight shift in the origin of the coordinate system, while $(q,\,\alpha)$ should be changed to 
\begin{equation}
 q = {7.5 \over {1000+7.5}} \times 10^{-3}=7.444 \times 10^{-3},\quad \alpha_0 = 360^\circ-\alpha = 86.37^\circ.
\end{equation}

%\include{tab1}

%\documentclass[10pt,preprint]{aastex}
%\usepackage{lscape}
%\usepackage{rotating}
%\begin{document}
%\headsep = 105pt
\begin{table*}
\caption{\label{tab:para}\sc Lens Parameters of the Examples}

\vskip 1em
(1) Example of the continuous external shear degeneracy shown in Figure \ref{fig:4-fold}.  
\vskip 1em
\begin{tabular}{@{\extracolsep{0pt}}cccccccccccccccc} 
\hline \hline
Fig. &$u_0$& {$m_2$} & {$m_3$} & {$s_{12}$} & {$s_{13}$} & {$\varphi$} & {$\alpha_0$} & Note\\ 
 & &$\times10^3$&$\times10^3$& & &(deg)&(deg) & \\ \hline \hline
   \ref{fig:4-fold}(a) &0.01&1.00&1.00&1.20&1.25&60.00&150.00 & wide-wide \\ \hline
   \ref{fig:4-fold}(b) &0.01&1.00&1.00&1.20&0.80&60.00&150.00 & wide-close \\ \hline
   \ref{fig:4-fold}(c) &0.01&1.00&1.00&0.83&0.80&60.00&150.00 & close-close \\ \hline
   \ref{fig:4-fold}(d) &0.01&1.00&1.00&0.83&1.25&60.00&150.00 & close-wide \\ \hline
\end{tabular}

\vskip 1em
(2) Examples of the continuous external shear degeneracy shown in Figure \ref{fig:triple}.  
\vskip 1em
\begin{tabular}{@{\extracolsep{0pt}}cccccccccccccccc} 
\hline \hline
Fig. &$u_0$& {$m_2$} & {$m_3$} & {$s_{12}$} & {$s_{13}$} & {$\varphi$} & {$\theta$} & {$\alpha_0$} & {$m_2'$} & {$m_3'$} & {$s_{12}'$} & {$s_{13}'$} & {$\varphi'$} & {$\theta'$} & {$\alpha_0'$} \\ 
 & &$\times10^3$&$\times10^3$& & &(deg)&(deg)&(deg)&$\times10^3$&$\times10^3$& & &(deg)&(deg)&(deg) \\\hline \hline
   \ref{fig:triple}(a)(b)&-0.0010&4.000&6.000&3.300&3.900&30.00&0.0&120.00&3.023&8.080&3.298&4.256&25.00&0.0&120.00 \\  \hline
   \ref{fig:triple}(c)(d)&-0.0015&4.000&6.000&3.300&3.900&45.00&0.0&120.00&4.002&9.391&3.666&4.842&40.00&0.0&120.00 \\ \hline
   \ref{fig:triple}(e)(f)&0.0015&4.000&6.000&3.300&3.900&150.00&0.0&165.00&4.993&5.642&3.611&3.819&149.00&0.0&165.00 \\ \hline
\end{tabular}

\vskip 1em
(3) Examples of the double-triple lens degeneracy from double-lens to
triple-lens shown in Figure \ref{fig:double2}.
\vskip 1em
\begin{tabular}{@{\extracolsep{0pt}}cccccccccccccccc} 
\hline \hline
Fig. &$u_0$& {$m_2$}& {$s$} & {$\theta$} & {$\alpha_0$} & {$m_2'$} & {$m_3'$} & {$s_{12}'$} & {$s_{13}'$} & {$\varphi'$} & {$\theta'$} & {$\alpha_0'$} \\ 
 & &$\times10^3$& &(deg)&(deg)&$\times10^3$&$\times10^3$& & &(deg)&(deg)&(deg) \\\hline \hline
   \ref{fig:double2}(a)(b)&-0.0010&8.000&4.000&5.0&115.00&5.123&2.502&3.870&3.796&15.00&0.0&120.00 \\  \hline
   \ref{fig:double2}(c)(d)&0.0282&7.444&1.306&2.0&86.37&2.475&4.955&1.303&1.304&3.00&0.0&88.37 \\ \hline
   \ref{fig:double2}(e)(f)&0.0282&7.444&1.306&3.0&86.37&2.959&4.445&1.299&1.301&5.00&0.0&89.37 \\ \hline
\end{tabular}

\vskip 1em
(4) Examples of the double-triple lens degeneracy from triple-lens to
double-lens shown in Figure \ref{fig:double1}.
\vskip 1em
\begin{tabular}{@{\extracolsep{0pt}}cccccccccccccccc} 
\hline \hline
Fig. &$u_0$& {$m_2$} & {$m_3$} & {$s_{12}$} & {$s_{13}$} & {$\varphi$} & {$\theta$} & {$\alpha_0$} & {$m_2'$} & {$s'$} & {$\theta'$} & {$\alpha_0'$} \\ 
 & &$\times10^3$&$\times10^3$& & &(deg)&(deg)&(deg)&$\times10^3$& &(deg)&(deg) \\\hline \hline
   \ref{fig:double1}(a)(b)&-0.0010&4.000&6.000&3.300&3.900&2.00&0.0&120.00&9.805&3.589&0.0&120.00 \\  \hline
   \ref{fig:double1}(c)(d)&-0.0010&4.000&6.000&3.300&3.900&0.00&0.0&120.00&9.796&3.586&0.0&120.00 \\  \hline
\end{tabular}

\end{table*}
%\end{document}

\label{lastpage}
\end{document}